# Switch and template pattern formation in a discrete reaction-diffusion system inspired by the *Drosophila* eye

Matthew W. Pennington[1] and David K. Lubensky[2]

1-The University of Michigan – Ann Arbor, Biophysics Program, 450 Church St., Ann Arbor, MI, 48109

2-The University of Michigan – Ann Arbor, Physics Department, 450 Church St., Ann Arbor, MI, 48109

PACS:

## ABSTRACT

We examine a spatially discrete reaction diffusion model based on the interactions that create a periodic pattern in the *Drosophila* eye imaginal disc. This model is capable of generating a regular hexagonal pattern of gene expression behind a moving front, as observed in the fly system. In order to better understand the novel "switch and template" mechanism behind this pattern formation, we present here a detailed study of the model's behavior in one dimension, using a combination of analytic methods and numerical searches of parameter space. We find that patterns are created robustly provided that there is an appropriate separation of timescales and that self-activation is sufficiently strong, and we derive expressions in this limit for the front speed and the pattern wavelength. Moving fronts in pattern-forming systems near an initial linear instability generically select a unique pattern, but our model operates in a strongly nonlinear regime where the final pattern depends on the initial conditions as well as on parameter values. Our work highlights the important role that cellularization and cell-autonomous feedback can play in biological pattern formation.

## 1. Introduction

Scientists have long been fascinated by the ability of biological systems to organize themselves into complex structures. The appearance of periodic patterns of gene expression and cell fate during animal development, in particular, has been studied for some time; starting with the famous work of Turing [1], a number of elegant mechanisms have been suggested that might underlie such pattern formation [2-5]. Only relatively recently, however, has it become possible to subject these ideas to direct experimental tests and to reconcile them with descriptions more firmly grounded in known genetic and molecular interactions. While confirming the utility of many of the classic proposals, this ongoing work has also made clear that further insights will be required to explain the richness and robustness of pattern formation during development [6-8]. In one example of the newer generation of models informed by detailed genetic studies, we argue in a separate communication that a novel *switch and template* mechanism is responsible for the hexagonal pattern of gene expression seen in the eye imaginal disc of the fruit fly *Drosophila melanogaster* [9]. Here, we give a detailed analysis of this new mode of pattern formation in its simplest, one-dimensional form.

The *Drosophila* eye imaginal disc is a monolayer epithelium—that is, a roughly two-dimensional sheet of cells—found in the fly larva and destined to develop into the adult fly's retina [10]. During the larval stage, a moving front of differentiation sweeps across the disc, leaving in its wake a regular lattice of single cells expressing the gene *atonal* (*ato*) and fated to become R8 photoreceptors [11-13]. These R8 cells then induce surrounding cells to attain other photoreceptor and support cell fates. The support cells surrounding a given R8 in turn secrete one of the ~750 lenses on the surface of the compound eye, which are arranged in a startlingly regular two-dimensional hexagonal packing (Fig. 1). The ordered packing of the lenses thus reflects the original pattern of *ato* expression. A large body of genetic experiments gives a qualitative picture of the regulatory network responsible for creating this expression pattern, making the eye disc an excellent model system in which to study biological pattern formation.

In this paper, we consider a model abstracted from the experimentally determined interactions controlling *ato* expression. This model distills the regulatory network down to its essential features by lumping superficially redundant genes into single dynamical variables, each of which can be thought of as representing a given sort of regulatory feedback: cell-autonomous auto-activation (i.e. activation of gene expression in a given cell by high concentrations of the same gene's products within that cell, without any cell-to-cell communication); short-ranged, but cell-non-autonomous, inhibition; and longer-ranged activation. Together, these interactions are sufficient to generate a stable, stationary pattern behind a moving front, as seen in the fly eye disc. In the one-dimensional case that is our focus here, this pattern takes the form of single, regularly spaced cells with high *ato* expression separated by a number $n \geq 1$ of cells with negligible *ato* levels. (In two dimensions the same model yields the observed hexagonal pattern and further makes testable predictions about the physiological patterning process in flies;

these predictions, and their experimental confirmation, are discussed elsewhere [9].) The pattern is generated through the interaction of a bistable switch created by the cell-autonomous positive feedback with a spatially-varying template of diffusible inhibitor produced by cells behind the front. As the front—driven by the long-ranged activator—progresses, individual cells at the leading edge are induced to flip from the low to the high *ato* state. These cells then inhibit *ato* expression in their neighbors, creating a space between successive high *ato* cells whose size is determined by the range of the inhibitory signal.

Unlike many standard patterning scenarios, this switch and template mechanism does not involve any Turing-like instability [5], and the final pattern is not related in any simple way to a bifurcation of an initially uniform state. In particular, in contrast to patterns that can be described by an amplitude equation [14, 15], the pattern that appears behind the front in our model can depend on the initial conditions. This final pattern has some similarities with those found in other bistable activator-inhibitor systems [16-18], but differs from them in that it was selected by a moving front and that *ato*'s strictly cell-autonomous self-activation greatly increases the variety of allowed patterns. Our model thus suggests a new, robust route to pattern formation in biological systems.

In the remainder of this paper, we first give an overview of the biology of the eye imaginal disc (Sec. 2) and then introduce our simplified mathematical model of pattern formation in this system (Sec. 3). In Sec. 4, we construct front solutions to our model in the limit that front propagation is much slower than the dynamics of *ato* expression and inhibitor secretion. In this limit, cells flip from low to high *ato* concentration almost instantaneously on the timescale of the long-ranged activator, and it is thus possible to calculate the activator profile created by a given pattern growing with a given speed; similarly, given an activator field, one can determine which cells will flip and when. The full solution is then found by self-consistently matching the behavior on these two different scales. Sec. 5 compares these predicted solutions to the results of simulating the full model for 640,000 randomly selected parameter sets and finds generally good agreement. The parameter scan shows that the behavior of the model is predictable and extremely robust to parameter variation, as long as the assumptions informing our analytic understanding are met. Finally, in Sec. 6 we compare our picture to other models of pattern formation in biology and discuss some broader implications of our results.

## 2. Biological background

Like all insects, *Drosophila melanogaster* has a compound eye composed of about 750 facets called ommatidia [19]. Each ommatidium in the adult eye is centered on a suite of 8 photoreceptor neurons (R1-8) and comprises a total of 20 cells [20]. The ommatidia are remarkable for their identical appearance and for the fact that they are packed into a perfectly crystalline hexagonal array in the adult eye (Fig. 1). These clusters of cells are not clonally derived, but begin differentiating from the unpatterned epithelium of the eye imaginal disc during the third instar period of larval development [21]. Ommatidia are founded by single cells expressing the gene *atonal* which will eventually become R8 photoreceptors [22, 23]. These cells are specified and begin differentiating at the same time that a front of physical distortion (the morphogenetic furrow, MF) moves across the epithelium from posterior to anterior. In front of the MF the cells are unpatterned, while immediately behind the MF one finds a characteristic hexagonal pattern of single cells expressing *ato* against a background of undifferentiated cells. This process has been reviewed by several authors [11, 13, 24]. In subsequent steps, each R8 cell interacts with the surrounding epithelium, inducing adjacent uncommitted cells to differentiate into the other neurons and support cells observed in the adult eye [22, 23, 25-30].

The progress of the MF is driven by the morphogen Hedgehog (Hh) secreted by the differentiating neurons behind the MF, and by the secreted factor Decapentaplegic (Dpp) expressed in the MF itself, both of which activate *ato* expression [31-33]. *ato* encodes a basic helix-loop-helix (bHLH) transcription factor and is the characteristic proneural gene for R8 specification [22, 23]. Its expression is initially diffuse, but is refined to single (future R8) cells as the MF passes [24, 34-36]. As *ato* expression becomes confined to a subset of cells, Ato also activates the zinc finger transcription factor Senseless (Sens), which in turn further activates *ato* expression and which continues to be expressed in R8 cells into adulthood [30, 37-39].

Because they appear at a moving front, the columns of R8 cells are specified sequentially (By convention, lines of cells running parallel to the MF are called columns, and those running perpendicular are called rows.). The positions of R8 cells in successive columns are found to be strongly correlated—each column is staggered along its long axis by one-half a row spacing, producing a hexagonal packing of R8s. It thus seems reasonable that each column might be specified using the previous one as a template, and several authors have suggested that inhibitory signals might serve to carry the needed information from one column to the next [24]. The idea that each *ato*- expressing cell is able to repress its neighbors, preventing or suppressing their *ato* production, is termed lateral inhibition. The molecular mechanism of this inhibition in eye discs is not known in detail, though the Notch (N) receptor is certainly integral to it, as is the Notch ligand Delta (Dl) [40-43]. Loss-of-function (LOF) of either of these genes results in an overpopulation of R8 photoreceptors [27, 36]. There are also other genes involved in the patterning process with more subtle phenotypes, among them *scabrous* [44, 45].

Fig. 2A summarizes the genetic interactions just described. Together, they conspire to create a moving front behind which single cells expressing *ato* appear in a regular pattern. With this genetic network in place, we now turn our attention to the construction of a model that captures its essential features.

## 3. The model

The network diagram of Fig. 2A incorporates three distinct sorts of feedback loops, mediated by secreted activators (Hh and Dpp), by cell-non-autonomous inhibitors (Dl, Sca, and others), and by strictly cell-autonomous interactions (direct Ato self-regulation and positive feedback through Sens). Presumably, there are functional reasons that the fly has a regulatory network that includes more than one representative of each sort of interaction. In making a first attempt to understand the basic pattern formation mechanism in the eye imaginal disc, however, it seems reasonable to elide these distinctions and to consider a model with only three variables, each representative of one of the three types of feedback. Indeed, such a simplified model, summarized in Fig. 2B, can capture many features of R8 patterning. The variable $a$ plays roughly the role of

*atonal* and directly activates its own expression, while $h$ and $u$ provide, respectively, non-autonomous activation and non-autonomous inhibition. (Elsewhere, we have considered a model with a fourth variable reminiscent of *sens*; the delay in the positive feedback that is thus introduced is important for accurately recapitulating all of the stages in the refinement of *ato* expression to a single cell, but it has little effect on the ability of the system to form a pattern [9].) Because of the central role played by cell-autonomous interactions in the eye disc, it is important that any model respect the discrete nature of the cells that make up the epithelium. Our model thus takes the form of a set of coupled lattice differential equations, with each lattice site representing a single cell. After non-dimensionalization, the governing equations in any dimension take the form

$$\frac{\partial a_x}{\partial t} = f_{n_a}(\tfrac{a_x}{A_a}) - a_x + G \cdot g_{m_h, m_u}(\tfrac{h_x}{H}, \tfrac{u_x}{U})$$

$$\tau_h \frac{\partial h_x}{\partial t} = f_{n_h}\left(\tfrac{a_x}{A_h}\right) - h_x + D_h \Delta h \qquad (1)$$

$$\tau_u \frac{\partial u_x}{\partial t} = f_{n_u}\left(\tfrac{a_x}{A_u}\right) - u_x + D_u \Delta u$$

Here, the subscript $x$ indexes the lattice site, and $\Delta$ is the lattice Laplacian operator, which is dependent on lattice geometry. Each variable has been rescaled so that its natural scale is of order unity. We have chosen to non-dimensionalize time by the decay rate of $a$; $\tau_h$ and $\tau_u$ give the decay times of $h$ and $u$ in these units. The source term $f_{n_\alpha}\left(\tfrac{a}{A_\alpha}\right)$ in each equation is a dimensionless function with $0 \leq f_{n_\alpha} \leq 1$ for $0 \leq a < \infty$. This restriction reflects the fundamental limits to the rate of production of any biomolecule. For simplicity, we take $f_{n_\alpha}$ to have the sigmoidal form

$$f_{n_\alpha}\left(\tfrac{a}{A_\alpha}\right) = \frac{a^{n_\alpha}}{A_\alpha^{n_\alpha} + a^{n_\alpha}}. \qquad (2)$$

This introduces the three dimensionless parameters $A_a$, $A_h$, and $A_u$ that characterize the scale at which $a$ activates production of itself, of $h$, and of $u$, and the three Hill coefficients $n_a$, $n_h$, and $n_u$. $D_h$ and $D_u$ are diffusion coefficients. We non-dimensionalize length by requiring that the lattice spacing be of order 1.

There are two terms in the equation for $\frac{\partial a}{\partial t}$, beyond its linear decay. They can be thought of as reflecting the presence of two enhancers at the *ato* gene, one responsible for auto-activation and the other for responding to the diffusible activators Hh and Dpp [9]. We assume here that inhibition acts primarily on this latter enhancer. The corresponding term in Eq. (1) has a maximum strength (relative to self-activation) $G$; the function $g$ varies between 0 and 1. A Hill-like functional form for this interaction offers, once more, the desired behavior in a simple package. Based on the fact that negative signaling through a pathway involving Delta, Notch, and Scabrous seems able to dominate any quantity of hedgehog-mediated enhancement, we used a non-competitive model for the interaction of these signals:

$$g_{m_h, m_u}\left(\frac{h}{H}, \frac{u}{U}\right) = \frac{(h/H)^{m_h}}{1+(h/H)^{m_h}} \left( \frac{1}{1+(u/U)^{m_u}} \right). \qquad (3)$$

The remaining model parameters ($H$, $U$, $m_h$, and $m_u$) are thus defined as the scales at which $h$ and $u$ become effective and the associated Hill coefficients.

Eq. (1) as written can describe a model in any dimension, but in this paper our primary interest is the one-dimensional version. On a regular 1D grid with nearest-neighbor interactions, the system reduces to a tridiagonal system of ODEs, where the integer-valued variable, $x$ indexes cells by their location in the grid. Exhibiting this spatial dependence explicitly and substituting the Hill-like forms for the functions $f$ and $g$, we arrive at the basic system of equations that we will study for the remainder of this paper:

$$\frac{\partial a_x}{\partial t} = \left(\frac{a_x^{n_a}}{a_x^{n_a} + A_a^{n_a}}\right) - a_x + G\left(\frac{\left(\frac{h_x}{H}\right)^{m_h}}{\left(\frac{h_x}{H}\right)^{m_h}+1}\right)\left(\frac{1}{\left(\frac{u_x}{U}\right)^{m_u}+1}\right)$$

$$\tau_h \frac{\partial h_x}{\partial t} = \left(\frac{a_x^{n_h}}{a_x^{n_h} + A_h^{n_h}}\right) - h_x + D_h\left(h_{x+1} - 2h_x + h_{x-1}\right) \qquad (4)$$

$$\tau_u \frac{\partial u_x}{\partial t} = \left(\frac{a_x^{n_u}}{a_x^{n_u} + A_u^{n_u}}\right) - u_x + D_u\left(u_{x+1} - 2u_x + u_{x-1}\right)$$

Our main interest in this work is the formation of patterns in which only single, isolated cells retain high $a$ levels. This requires that these cells be able either to keep their neighbors from becoming activated, inhibiting them before their concentrations begin to rise, or to force them back down after their $a$ levels have begun to move upwards. The former scenario turns out to hold for parameter sets we have found that consistently form patterns. In this case, levels of the inhibitor $u$ must respond quickly to changes in $a$, and we thus expect $\tau_u$ to be small. It is then natural to simplify our analysis by taking the limit $\tau_u \to 0$, replacing Eq. (4) by

$$\frac{\partial a_x}{\partial t} = \left( \frac{a_x^{n_a}}{a_x^{n_a} + A_a^{n_a}} \right) - a_x + G \left( \frac{\left(\frac{h_x}{H}\right)^{m_h}}{\left(\frac{h_x}{H}\right)^{m_h} + 1} \right) \left( \frac{1}{\left(\frac{u_x}{U}\right)^{m_u} + 1} \right)$$

$$\tau_h \frac{\partial h_x}{\partial t} = \left( \frac{a_x^{n_h}}{a_x^{n_h} + A_h^{n_h}} \right) - h_x + D_h \left( h_{x+1} - 2h_x + h_{x-1} \right) \qquad \text{(5a, 5b, 5c)}$$

$$0 = \left( \frac{a_x^{n_u}}{a_x^{n_u} + A_u^{n_u}} \right) - u_x + D_u \left( u_{x+1} - 2u_x + u_{x-1} \right)$$

That Eq. (5c) is linear in $u$ and non-singular makes this a relatively benign change. In effect, in this limit, $a$ can be regarded as having a local, cell-autonomous effect which may be activating or inhibitory, and a non-local effect that is always inhibitory. The substitution of Eq. (5) for Eq. (4) further prohibits cell-autonomous oscillations that could be spawned from the relaxation-oscillator-like structure of the activator-inhibitor system.

## 4. Front solutions

In this section, we construct solutions to Eq. (5) in the limit $\tau_h \gg 1$ and $D_h \gg 1$ that have the form of a front that moves with constant velocity and leaves a regular periodic pattern in its wake. This limit is consistent with the properties of the activators present in the fly system: Hh, in particular, diffuses forward from differentiating cells behind the MF and thus must be quite long-ranged. In order for front velocity to remain of order unity, $\tau_h$ must then also be large. Our strategy makes use of this separation of timescales to determine the behavior of $a$ and $u$ on short length scales given an imposed $h_x(t)$ and to find the $h$ front created by a lattice of sources (i.e. cells with high $a$) that adds a new cell at regular intervals. These solutions on two different scales are then matched self-consistently to arrive at the full front solution. Fig. 3 illustrates some observed behaviors of our one-dimensional system, while Fig. 4 gives a spatiotemporal portrait of the regular, patterning solutions that are our primary interest here.

### 4.1. Cell-autonomous behavior

Since our model consists of coupled ODEs on a lattice, we can ask about the behavior of a single, isolated cell or lattice site and separate the influences of cell-autonomous and non-autonomous interactions. Towards this end we first solve for the steady-state $u$ field due to a source of strength $s$ at $x = 0$ in an infinite 1D grid with zero boundary conditions at $x = \pm\infty$,

$$0 = s\delta_x - u_x + D_u \left( u_{x-1} - 2u_x + u_{x+1} \right)$$
$$u_{\pm\infty} = 0 \tag{6}$$

The solution is elementary and is given by

$$u_x = c_0 \lambda^{|x|}$$
$$\lambda = \frac{1 + 2D_u - \sqrt{1 + 4D_u}}{2D_u} \tag{7}$$
$$c_0 = s \cdot \frac{-1 + \sqrt{1 + 4D_u}}{1 + 4D_u - \sqrt{1 + 4D_u}}$$

The quantity $c_0$ is the contribution of a cell producing $u$ to the local amount of that substance at the same cell. If $D_u$ approaches zero, nothing diffuses, all of the $u$ remains local, and $c_0 \to s$. Conversely, if $D_u$ is large, most of the substance diffuses away, and its local influence tends to zero. Because inhibition acts over relatively short distances, on the scale of pattern wavelength, we expect $D_u$ to be nearer the first extreme. In contrast, we have already argued that $D_h \gg 1$. Examining a single cell, then, we must separately consider $u$ produced locally, by that same cell, and exogenously, whereas $h$ is mainly exogenous.

Neglecting autoinhibition, the amount of $a$ at a lattice site can be bistable through autoactivation (Fig. 5A). Here, we focus primarily on the simple case in which cells starting at low $a$ can switch to high $a$ or remain at low values of $a$, but a cell with high $a$ cannot go back down. Increased $h$ can flip a cell to the high state as long as $u$ does not block $h$'s effects; with $g_{m_h, m_u}\left(\frac{h}{H}, \frac{u}{U}\right)$ (which, in a slight abuse of notation, we will sometimes abbreviate as $g(h,u)$) viewed as a fixed, externally imposed bifurcation parameter, the low steady state collides with the unstable saddle in a saddle-node bifurcation. For this change to be irreversible, the complementary bifurcation (the one which would lead to the disappearance of the high steady state with decreasing $h$) must not be accessible, even at maximal inhibition or zero activation ($u \gg U$, or $h=0$). These cases are illustrated in Fig. 5B. Restricting the bistable switch in this way prevents the transient formation of a high-amplitude pattern. Restrictions on $A_a$ sufficient for the existence of bistability for some non-negative value of $g(h,u)$, and bistability at $g(h,u)=0$ are, respectively,

$$A_a < \frac{1}{4}\left(\frac{n_a^2-1}{n_a}\right)\left(\frac{n_a+1}{n_a-1}\right)^{1/n_a}, \tag{8}$$
$$n_a > 1$$

and

$$A_a < \frac{(n_a-1)^{\frac{n_a-1}{n_a}}}{n_a}. \tag{9}$$
$$n_a > 1$$

Evaluated for $n_a = 4$, these require $A_a < 1.065$ for bistability and $A_a < .569$ for irreversible bistability. As already mentioned, we are primarily interested in the more stringent requirement.

The next step in treating a single cell is to include the $u$ that was produced by the cell itself, which we call the self-$u$. With $\tau_u = 0$, a cell with activator concentration $a$ creates a $u$ concentration $c_0$ locally, as given by Eq. (7), with $s = f_{n_u}\left(\frac{a}{A_u}\right)$. Below, we have introduced the variable $u_{ns}$ (meaning $u$-non-self) to represent inhibitor produced elsewhere that has diffused to the current location. With $h$ and $u_{ns}$ still viewed as fixed parameters, the $a$ concentration in a single cell then obeys

$$\partial_t a = \frac{a^{n_a}}{A_a^{n_a} + a^{n_a}} - a + G \cdot \bar{g}(h)\left[1 + \left(\frac{u_{ns} + c_o\left(\frac{a^{n_u}}{A_u^{n_u}+a^{n_u}}\right)}{U}\right)^{m_u}\right]^{-1}. \tag{10}$$

Here, the function $\bar{g}(h) = h^{m_h}/(H^{m_h} + h^{m_h})$ separately keeps track of the activating contributions to $g_{m_h,m_u}\left(\frac{h}{H}, \frac{u}{U}\right)$.

Figs. 5C and 6 illustrate the behavior of an isolated cell with autoinhibition governed by Eq. (10). If $A_u$ or $U$ is too small, autoinhibition will be strong enough to completely abolish the lower saddle-node bifurcation, and a cell that is initially in the low state will remain there forever. Similarly, but physically more interesting, since it depends on a non-autonomous quantity, enough $u_{ns}$ can make the high steady state completely inaccessible from the low steady state. That is, there is a threshold value, $u_{threshold}$ for $u_{ns}$ above which $g(h,u) < g_c$, where $g_c$ is the critical value to flip the switch, for any $h$. For typical parameter values, the high steady state itself is nearly invariant over the range of $h$ because autoinhibition effectively blocks all activation through the 3' enhancer when $a$ is high. This effect will be important when we consider templating in sec. 4.2.2.

The dynamics by which a cell can make the traverse from low to high steady state is also readily understood. The clearest feature of Fig. 6B is that there is a region of $a$ dynamics where $a$ is high enough to cause significant autoinhibition, and its approach to the high steady state is thus nearly independent of $\bar{g}(h)$, and by extension of $u_{ns}$. In this range $a$'s dynamics are instead governed almost exclusively by its intrinsic autoactivation timescale. The region of $a$ dynamics dependent on $h$ shows the potential for a bottleneck if $\bar{g}(h)$ exceeds the bifurcation value very slowly. If a cell is stuck in this bottleneck, it is still susceptible to repression by $u_{ns}$. This is rarely an issue in a 1D system, in which a front propagating with velocity $v$ produces a delay of order $\frac{1}{v}$ between neighboring cells, but is potentially important in 2D, where no delay need exist for cells adjacent in a direction perpendicular to the direction of front propagation.

In summary, each cell can act as a bistable switch, with the slow variable $g(h,u)$ or $\bar{g}(h)$ as bifurcation parameter, and exogenous inhibitor $u_{ns}$ tuning the switches sensitivity; adding the effect of endogenously generated $u$ does little to change these basic properties. The dynamics of activation are usually dominated by properties intrinsic to the cell.

### 4.2. Propagating solutions

Inspired by the behavior observed in imaginal discs, we are interested in solutions to Eq. (5) propagating with constant speed that produce a regular pattern of isolated *active* (high $a$) cells separated by an integral number of *inactive* (low $a$) cells (one-up-integer-down patterning, or OUID). Such solutions really involve two distinct processes occurring on different time and length scales. One is a templating process by which lateral inhibition selects a pattern; the second is the process by which the pattern is pushed forward by the action of $h$. Given the behavior of the template, specifically the end pattern and the rate at which it is produced, we can calculate $h$ at any point (sec. 4.2.1). Similarly, $h$ as a function of *x* and *t*, we can calculate the pattern produced (sec. 4.2.2). One can then look for self-consistent solutions where the $h$ produced by a pattern of activated cells interacts with the template produced by those cells in such a way that the original pattern is extended (sec. 4.2.3). A solution of this sort, if stable, should represent an observable asymptotic long-time behavior of the model.

### 4.2.1. The h field due to a periodic pattern

Above, in Eq.(7), we quoted the steady-state distribution due to a point source of a substance diffusing on a 1D lattice. To deal with the dynamics of the propagating $h$ front, we need more detailed information. We would like to solve the problem

$$\begin{aligned} \tau_h \partial_t h_x &= s_x(t) - h_x + D_h(h_{x-1} - 2h_x + h_{x+1}) \\ s_x(t) &= \bar{s}(vt - x)\delta_{x, q \cdot j} \\ j &\in \mathbb{Z} \end{aligned} \qquad (11)$$

Here $\bar{s}(t)$ gives the stereotyped dynamics of a cell being activated. $s_x(t)$ thus corresponds to a pattern with integer period $q$ growing with speed $v$. In the limit $\tau_h \gg 1$, cells flip from low to high $a$ almost instantaneously on $h$'s timescale. The source term $s_x(t)$ correspondingly jumps from a value near zero (for typical parameters) to a value determined by the high steady state of $a$, which we call $s_0$. Because of our choice of nondimensionalization and the strong effect of auto-inhibition in active cells, $s_0$ is generally very near 1. The explicit time dependence of $s$ then becomes

$$s_x(t) = s_0 \Theta(vt - x)\delta_{x, q \cdot j}, \qquad (12)$$

where $\Theta$ is the Heaviside step function. The impulse response of the differential equation system (11) is known exactly. The general solution is then the sum over all cells of an integral over time, where the integrand is the product of the source strength and an exponentially modulated associated Bessel function of the 1st kind, $I$:

$$h_x(t) = \frac{1}{\tau_h} \sum_{x'=-\infty}^{\infty} \int_{-\infty}^{t} s_{x'}(t') e^{-(t-t')-2D_h(t-t')/\tau_h} I_{|x-x'|}\left[\frac{2D_h}{\tau_h}(t-t')\right] dt'. \quad (13)$$

Applying the idealized form of $s_x(t)$ from Eq. (12) leads to the simplified expression:

$$h_x(t) = \frac{s_0}{\tau_h} \sum_{x'<vt, x'=q\cdot j} \int_0^{t-x'/v} e^{-(1+2D_h)t'/\tau_h} I_{|x-x'|}\left(\frac{2D_h}{\tau_h}t'\right) dt'. \quad (14)$$

To better understand this formula, it helps to look at the analogous continuum problem, where the source term is not patterned and $x$ is a continuous spatial variable, namely,

$$\tau_h \partial_t h(x,t) = s_0 \Theta(vt-x) - h + D_h \partial_x^2 h. \quad (15)$$

This problem can be solved exactly by transforming into reference frame moving at a speed $v$. Applying appropriate boundary conditions yields the following result:

$$h(x,t) = s_0 \begin{cases} 1 - \left(\frac{v\tau_h + c_1}{2c_1}\right) \exp\left[\frac{-v\tau_h + c_1}{2D_h}(x-vt)\right], x < vt \\ \left(\frac{-v\tau_h + c_1}{2c_1}\right) \exp\left[\frac{-v\tau_h - c_1}{2D_h}(x-vt)\right], x \geq vt \end{cases} \quad (16)$$

$$c_1 = \sqrt{v^2 \tau_h^2 + 4D_h}$$

If activation is instantaneous, then the amount of $h$ at $x = vt$, the point where $h$ production has just been activated, should be exactly the value required to trigger the bistable switch, as discussed above in Sec. 4.1. The $h$ at this point decreases monotonically as $v$ increases, and knowing the critical value of $h$ for activation and the source strength $s_0$ uniquely determines the velocity $v$, which was previously arbitrary.

If one attempts to treat a continuum system with a spatially periodic (rather than constant) source term, static in the lab frame, by transforming into a moving reference frame, a point at constant $z = x - vt$ does not approach a steady state for any $v \neq 0$, but instead oscillates, with average $h$ given by Eq. (16) for an appropriate choice of $s_0$. The amplitude of the deviations of $h$ from the ideal, unpatterned case will be relatively small if the largest spatial scale of the source pattern is small compared to the smallest spatial scale in the propagating front. This spatial scale is equal to $\sqrt{D_h}$ at $v = 0$, which remains a good estimate for most parameter sets as typically $D_h \gg \tau_h^2 v^2$. For the reference parameter value $D_h = 640$, $\sqrt{D_h} \approx 25.3$ and patterns with period 5 (roughly the scale that the fly eye disc leads us to be most interested in) are effectively averaged over.

Significant work has been done on the discrete version of this problem, though never with a patterned template [46-50]. The basic result we rely on here is that any deviations from the continuum behavior tend to become inconsequential for $\sqrt{D_h} \gg 1$. To the degree that this condition holds, we can approximate the exact discrete expression of Eq. (13) by sampling the continuum solution Eq. (16) at regular intervals: $h_x(t) \approx h(x,t)$. We use this approximation to give boundary conditions for simulations of propagating systems which are necessarily conducted on finite grids of cells but where we are interested in finding the model's asymptotic long-time behavior (Sec. 5 and Appendix A, below).

### 4.2.2. Template Formation in 1D

So far we have assumed the existence of a large-amplitude periodic pattern of $a$ which acts as a source for $h$. We now turn to the question of whether such a pattern will indeed form in response to a moving front of $h$ of the form of Eq. (16). We first discuss in detail a simple limiting case in which smoothly varying $h_x(t)$ is replaced by a step function and then develop analogous arguments for the more general case. Some further technical details are found in Appendix B.

With $u$ slaved to $a$, the total $u$ concentration at point $x$ follows directly from Eq. (7) and is

$$u_x = \sum_{x'} c_{0,x'} \lambda^{|x-x'|}$$
$$\lambda = \frac{1+2D_u - \sqrt{1+4D_u}}{2D_u} \qquad (17)$$
$$c_{0,x'} = \left(\frac{a_{x'}^{n_u}}{A_u^{n_u} + a_{x'}^{n_u}}\right) \cdot \frac{-1+\sqrt{1+4D_u}}{1+4D_u - \sqrt{1+4D_u}}$$

For any point in advance of a regularly patterned half-space of identical activated cells, $u_x$ is a geometric series that converges increasingly rapidly for small $D_u$. For a simple pattern with period $q$, so that active cells are found at $x' = 0, -q, -2q, -3q...$, with identical $u$ production at each (as will be the case if the $u$ production term $f_{n_a}(a_{x'}/A_u)$ is saturated or if $a_{x'}$ varies little among the active cells), the expression for $u$ at $x \geq 0$ is

$$u_x = c_0 \frac{\lambda^x}{1-\lambda^q}, \qquad (18)$$

where $c_0$ is given by $c_{0,x'}$ in Eq. 17 and is assumed to be the same for $x' = 0, -q, -2q, -3q...$. Here, and throughout this section, we assume that cells that have not been activated produce negligible $u$. In other words, we replace the Hill function $f_{n_u}\left(\frac{a}{A_u}\right)$ with a step function. Thus, at inactive cells only non-self $u$ is present, and $u = u_{ns}$, while at active cells $u = u_{ns} + c_0$.

In the section on cell-autonomous behavior, we noted that there is, in general, an amount of exogenous $u$ (call it $u_{threshold}$) that can absolutely prevent $a$ from leaving the low steady state, regardless of activation from $h$. Suppose first that the role of inhibition in this model is solely to put some cells above this threshold, so that when the $h$ front progresses, the next cell to be activated is just the first one it encounters with $u_x < u_{threshold}$. That is, the first cell that possibly can be activated by $h$ is the one that actually does come up.

A self-extending pattern in these conditions is subject to two inequalities which ensure that the next cell to come up is spaced a distance from the previous cell that is the same as the period $q$ of the existing pattern:

$$c_0 \frac{\lambda^q}{1-\lambda^q} < u_{threshold} < c_0 \frac{\lambda^{q-1}}{1-\lambda^q}. \tag{19}$$

For given constants $c_0$, $\lambda$, and $u_{threshold}$, there is no more than one integer $q$ that satisfies these conditions. We understand the physical meaning of the two cases (integer solution exists or does not) in terms of a one-dimensional map that gives $u$ in each newly activated cell in terms of the value of $u$ in the previous activated cell. Let the previously activated cell be located at $x=0$ with inhibitor concentration $\hat{u}_m$ immediately after activation and the newly activated cell have inhibitor concentration $\hat{u}_{m+1}$ immediately after its own activation, where $m$ indexes the active cells and plays the role of time in the map. If the newly activated cell is at spatial position $x' > 0$, then since the $u$ produced at all preceding cells is decaying with the same spatial dependence, we have a map of $\hat{u}_m$ onto $\hat{u}_{m+1}$, where $c_0$ accounts for the inhibitor produced by the newly activated cell itself:

$$\hat{u}_{m+1} = \hat{u}_m \lambda^{x'} + c_0. \tag{20}$$

For the cell at $x'$ to indeed be the next one activated (and thus the first one that can be activated), the preactivation $u$ must satisfy $u_{x'} = \hat{u}_m \lambda^{x'} < u_{threshold} < u_{x'-1} = \hat{u}_m \lambda^{x'-1}$. The full map giving the value of $u$ at the cell that is actually activated is thus the union of segments of the form of Eq. (20) for different $x'$ that obey these inequalities, and is consequently piecewise linear and discontinuous. All the linear segments have positive slope less than one (since $\lambda < 1$), and the inequalities are such that all the segments with

$x' > 1$ lie within a finite band of allowed $\hat{u}_{m+1}$, as shown in Fig. 7. Illustrations of these maps are shown in Fig. 7. This kind of map has been described by Jain and Banerjee [51]. There are two possibilities: the identity line passes either through a line segment, such that there is one point of intersection (i.e. one integer $x'$ satisfying Eq. (20) with $\hat{u}_{m+1} = \hat{u}_m$ [Fig. 7A]), or through a discontinuity between two line segments (no integer $x'$ satisfying Eq. (20) with $\hat{u}_{m+1} = \hat{u}_m$ [Fig. 7B]). The behavior encountered in going from the first situation to the second is called a discontinuous border collision bifurcation. At such a bifurcation, a single, stable, period-1 attractor is replaced by a stable limit cycle with period greater than one. Fig. 8 for illustrates a case with a period-3 solution. These solutions reach arbitrarily high periods, and are arranged in parameter space in a complex geometry.

Translating our results on the map back into the language of spatial patterns, we find that if one varies parameters continuously in such a way as to go from a stable OUID pattern of period $q$ to a period $q+1$ pattern—which, in terms of the map $\hat{u}_m \to \hat{u}_{m+1}$ correspond, respectively, to fixed points on the $q^{th}$ and $q+1^{th}$ line segments—one must pass through a region characterized by more complex high-period patterns. These patterns correspond to limit cycles in the map that oscillate between the $q^{th}$ and $q+1^{th}$ line segments, as illustrated in Figs. 7 and 8. The unit cell of the pattern thus consists of a set of single active cells separated by either $q-1$ or $q$ inactive cells, with the sequence of $q$ and $q-1$ dictated by the limit cycle in the map. Notably, in this simplified picture the patterning solution is globally attractive given any initial prepattern, and is unique up to an overall translation [51].

These results can be generalized for significantly relaxed assumptions. In the original model of Eqs. (4) and (5), $h$ can take on any number of intermediate values (whereas in the preceding paragraphs we have, in effect, taken it to be a step function in space), and the effect of $u$ is not necessarily negligible for $u < u_{threshold}$. To address the second concern first, we note that, in general, there is a critical value of $h$, $h_{crit}(u)$, that gives the bifurcation value where the low steady state of $a$ disappears. $u_{threshold}$ in the previous analysis is defined so that $\lim_{u \to u_{threshold}} h_{crit}(u) = \infty$. More generally, $h_{crit}(u_x)$ decreases monotonically as a function of increasing $x$ toward a finite, positive limiting value. Additionally, its second derivative in space (and discrete approximations thereto) is always positive. These general characteristics of $h_{crit}$ are dictated by the functional form of $u_x$ in our model.

To start dealing with the continuous variability of $h$ and its spatial structure, let us approximate the advancing front of activation with a linear function of $z = x - vt$ restricted to positive numbers, and with slope $-c_3$. This is reasonable when $\sqrt{D_h} \gg 1$ so that $h$ varies slowly in space compared to the scale of the pattern. We then have

$$h = \max[-c_3 z, 0]. \tag{21}$$

Unlike in our earlier treatment of a continuum model of $h$ production (Eqs. 15-16), here $z = 0$ does not necessarily correspond exactly to the point where an active cell first appears. As before, assume that there is a semi-infinite regular pattern on $x \leq 0$ and ask where the next cell is activated. As $t$ increases, the first cell where $h$ exceeds $h_{crit}(u_{x'})$ is, again, found at a point $x'$ governed by two inequalities (see also Appendix B),

$$h_{crit}(u_{x'+1}) - h_{crit}(u_{x'}) < c_3 < h_{crit}(u_{x'}) - h_{crit}(u_{x'-1}). \tag{22}$$

These are analogous to the simpler inequalities for $u$ of Eq. (19), and they determine the locations of the discontinuities in a similar map of $\hat{u}_m$ onto $\hat{u}_{m+1}$. One can show that the existence of a unique, globally attractive fixed point or limit cycle for the map is guaranteed by the positive second derivative of $h_{crit}(u_{x'})$, excepting the non-generic case of equality in one of the relationships in Eq. (22), which indicates simultaneous activation of two cells. Appendix B argues that this same qualitative behavior persists whenever $h_x(t)$ is a function only of $x-vt$ —which we have already pointed out is true to very good approximation for our system when $\sqrt{D_h}$ is large—and $u_x$ decays sufficiently fast that $u_{ns}$ at cells that are about to be activated comes almost entirely from the most recently activated cell.

We thus conclude that, for imposed $h_x(t)$ of the form (16) and $\tau_h, \sqrt{D_h} \gg 1$, there is a unique pattern that consists of single activated cells; the only exception occurs when $h_x(t)$ is too weak ever to activate any cells, even in the complete absence of inhibition. The resulting pattern has either a simple OUID form or a more complex periodic pattern consisting of single active cells separated by some admixture of two integer numbers of inactive cells. The former includes the possibility, when each cell produces a very small amount of inhibitor, of active cells separated by 0 inactive cells, in which case the resulting pattern of course consists entirely of cells with high $a$; we will see, however, that such a pattern more often arises because our assumption of a separation of timescales between $h$ and $a$ is violated.

### 4.2.3. Self-consistent solutions

Armed with these ideas for understanding front propagation and pattern templating, we now seek solutions to the full model where the $h$ front created by a pattern interacts with the template in such as way as to extend the same pattern indefinitely. If we restrict ourselves to OUID patterns, in addition to the parameters in the basic model, the solutions for $h_x(t)$ and the patterns found as solutions to the templating problem in the previous two sections are characterized by two quantities, the front speed $v$ and the pattern density $1/q$. Our goal is to find values for these two variables that allow us to match a short-scale templating solution to the large-scale $h$ front.

We first ask whether, for a given $q$, there is a velocity such that a source of the form Eq. (12) will produce an $h$ field $h_x(t)$ that in turn, through the mechanisms just described, creates a pattern of the same period $q$. It makes sense to treat front velocity as a continuous variable. Pattern density, on the other hand, is the fraction of cells with high $a$ in a regular pattern, and for OUID patterns, it is the inverse of an integer. For a fixed pattern density in the continuum limit ($s_0$ in Eq. (16)), the value of $h$ at the cusp of the front, $x = vt$, is a simple, monotonic function of velocity. If we have other information that dictates this $h$ concentration, we can solve for the velocity; in this case we set it equal to the critical value $h_{crit}$ needed to flip the bistable switch.

The discreteness of the system and the non-uniform pattern complicate this formulation only slightly. With the spatial pattern enforced externally, self-consistency demands that $h$ at the next point that must be activated to extend the pattern reach the triggering value at a particular time, extending the temporal pattern. This allows one parameter (the velocity $v_q$ appropriate to a specific pattern density, $1/q$) to be varied to meet this requirement, which we write as,

$$h_0(0) = h_{crit} = h_q\left(q/v_q\right), \qquad (23)$$

where we have taken the cell previously activated at $t=0$ to be at $x=0$ and the cell newly activated at $t=q/v_q$ to be at $x=q$, and where $h_{crit}$ is short for $h_{crit}(u_q)$ for $u$ generated by a semi-infinite pattern of period $q$.

This is a 1D root-finding problem of a monotonic function in a semi-infinite domain: as long as a solution $v_q$ exists, it is easy to find numerically by standard techniques (see Sec. 5). In the continuum, $h$ is capable of producing a moving front as long as $h_{crit} < s_0/2$. This is not the case in a discrete system, and propagation can fail at much lower $h_{crit}$ [46, 47]. Indeed, this propagation failure is a key prediction of our model. Since the amount of $h$ due to a static pattern increases monotonically with time to its steady state, the sufficient condition for the existence of a self-consistent velocity for a pattern of isolated active cells with period $q$ is that the steady state $h$ due to a semi-infinite pattern exceed the critical value at the next-to-be-activated point,

$$h_q(\infty) > h_{crit}(u_q) \Rightarrow v_q \in \mathbb{R} > 0. \tag{24}$$

Lower pattern densities obviously produce lower equilibrium values of $h$ at all cells. Since $h_{crit}$ decreases to a finite limit as $u \to 0$, there is always a minimum pattern density, $1/q_0$, that can be considered a candidate for a self-consistent period-$q$ OUID solution. Self-consistent velocities exist for all $q, 1 \leq q \leq q_0$, but not all of these choices of $q$ and $v_q$ correspond to actual solutions to the original model. We have imposed $q$ and calculated the velocity that pattern would produce, should it exist. Most of these patterns are pure speculation, since at the point when the cell that extends the pattern is activated, other cells may already have been activated, destroying the assumed pattern. We choose among these options, and thus solve for $q$, by requiring that the next cell to become activated is always the first one triggered by the previous pattern, a formulation of the fast-$a$ assumption. The idea here is that this cell will quickly become activated and suppress its near neighbors. This requirement is applied quantitatively by asking, for any given pattern expanding at its self-consistent rate, whether, at the moment of activation of the appropriate next cell, $h$ at any other cell is greater than that cell's $h_{crit}$. This question should be asked for every potential value of $q$. It is possible to pick parameters where the number of self consistent patterns is 0, 1, or more than one. Cases with one self-consistent solution and no self-consistent solutions correspond, respectively, to the cases discussed in the section on templating with single period-1 solutions or only high-period solutions. Parameter sets with more than a single self consistent solution correspond to cases where the change in the *shape* of the propagating $h$ front due to a change in the pattern density is enough to substantially change how $h$ interacts with the inhibitor template. Such cases are relatively rare, but we have verified numerically that they do indeed exist. It is also worth noting that, for the more general case where $h$'s spatial variation is taken into account, we have not explored more complex solutions corresponding to limit cycles in the one-dimensional map, though they can be found numerically in appropriate slices of parameter space.

## 5. Simulations

It remains to be seen how well the analytic results of section 4—which were, after all, obtained for a limiting case—can be applied to the original system of Eq. (5). We are particularly concerned with the degree to which this limiting behavior characterizes the actual system behavior over a large range of parameters, the robustness of the system to numerical variations in its parameters, and the adequacy of the assumptions that led to our detailed understanding of the system. By random parameter scanning over a large range, we have confirmed that, indeed, the behavior as predicted in sec. 4 is observed over a very large range of parameter space. Furthermore, the region of parameter space where the predicted behavior occurs is separated from the region where the primary unpredicted behavior occurs by a relatively straight line when projected to represent $a-h$ timescale separation, indicating the sufficiency of our assumptions. When viewing parameter sets far from this dividing line, we can conclude that the model is, in fact, extremely robust to parameter variations.

## 5.1. Parameter scan

In order to rest our analytic approach and to gain a fuller understanding of our model, we conducted a random parameter search in a region of parameter space known to contain at least some solutions that yielded behavior of interest. For the purposes of this scan we varied the concentration parameters $A_a$, $A_h$, $A_u$, $U$ and $H$; the operator-strength parameter, $G$; the diffusion constant parameters, $D_u$ and $D_h$; the Hill coefficient $m_h$; and the time-scale constant $\tau_h$. We centered the parameter search on a parameter set, $p_{ref}$, we knew to give results in 2D similar in appearance to the patterning observed in actual developing fly eyes [9].

$$p_{ref} = \begin{cases} A_a = .25 \\ G = 3.475 \\ H = .0193 \\ m_h = 8 \\ U = .00001048 \\ \tau_h = 371.65 \\ A_h = .75 \\ D_h = 640 \\ A_u = .9 \\ D_u = .16 \end{cases} \qquad (25)$$

We did not vary the Hill coefficients for functions of $a$ and $u$. These are summarized in $p_{static}$.

$$p_{static} = \begin{cases} n_a = 4 \\ n_u = 8 \\ n_h = 4 \\ m_u = 8 \end{cases} \qquad (26)$$

In scanning the parameter space we chose random parameter sets in which each parameter was chosen independently and either the parameter or its log was chosen to be uniformly distributed with an interval listed in Table I. We then made analytic predictions using the methods of Sec. 4 for the model's behavior for each choice of parameters, and compared these predictions to the result of directly integrating eq. (5). The limits of the sampled interval and the associated distribution used for each parameter are summarized in Table 1. We generated and tested 640,000 independent sets of parameters according to these rules.

The sampling limits for this work are necessarily a bit arbitrary; we took our target to be two orders of magnitude up and down from each reference value. This limit did not make sense for the variables, $A_a$, $A_u$, and $A_h$, as the model does little of interest if they exceed the high steady state of $a$, which is typically near 1. We limited the maximum value of $\tau_h$ for the practical reason that this plays a very direct role in how long an equation system must be integrated to examine its asymptotic behavior. That asymptotic behavior is expected to become independent of $\tau_h$ (up to an overall rescaling of time) for large enough $\tau_h$. Varying the Hill coefficient, $m_h$, is significantly different than varying the other parameters, in that it does not have an obvious ratiometric interpretation. Values of $m_h > 1$ represent sigmoidal curves for activation by $h$, whereas values of $m_h \leq 1$ represent the qualitatively different case where activation is most sensitive to changes in $h$ at $h = 0$.

We subjected every parameter set to the analyses described in Appendix A, and based on the analytic understanding we have outlined (Sec. 4), predicted whether we expect a patterning solution to exist, and what its speed and period should be if one does. We additionally integrated the model (Eq. (5)) for each parameter set on a 1D grid using random initial conditions chosen from an appropriate distribution, as well as initial conditions specifically meant to mimic the predicted asymptotic pattern-forming attractor. Using an automated pattern detection scheme, we compared the result of each integration both qualitatively and quantitatively to the predictions arrived at analytically.

## 5.2. Analyzing patterns

The first, qualitative, stage of looking at patterns involved classifying parameter sets amongst 5 basic types of behavior based on the results of simulations of the full set of governing Eq. (5): *Patterning* (Fig. 4A), *stalled* (Fig. 4B), *poorly-patterning*, *non-patterning* (Fig. 4C), and *impermanent* fronts. The first two cases, where a solution consisting of a self-extending periodic pattern of isolated active cells exists or where the system cannot sustain a moving front of any sort are addressed by our theory, and we expect predictions of behavior, period, and timing to be accurate in the limit that $a$ dynamics are much faster than $h$. The other cases are categories of behavior we observed in the course of running simulations that are not explained in detail by our theory, and represent the failure of our assumptions. Briefly, a poorly-patterning front consists of a solution in which an initial pattern leads to a propagating front resulting in some active cells and some inactive cells, but where these cells are not arranged in a periodic pattern with isolated active cells. Non-patterning fronts exist when an initial pattern leads to a propagating front where all the cells become active. Some instances of this behavior fall within the purview of our theory: we predict such a solution when an active cell does not produce enough inhibitor to prevent the activation of any of its neighbors. Impermanent fronts are any solutions in which a cell we determine to be active becomes inactive again at a later time. Such solutions violate one of our fundamental assumptions, namely the irreversibility of activation, and can be found only when $A_a$ is large enough to violate the inequality of Eq. (9).

It was our hope that the parameter sets showing behavior not predicted analytically would clearly be the result of the failure of one of the assumptions we made explicitly in our analysis, namely the separation of timescales between $a$ and $h$ dynamics and the irreversibility of $a$. In discussing patterns, we refer strictly to large-amplitude, long-lived patterns of $a$, as these are the characteristics of the biological system that we are modeling.

## 5.3. Results

### 5.3.1. Qualitative results

The results of our parameter scan were entirely consistent with the predictions of sec. 4; behaviors we did not find analytically began to appear only when the assumptions behind the analytics no longer held. The most interesting assumption, both because of the behavior observed during its failure and because it may not hold in all real-world conditions of interest, is the large separation of timescales between the advance of the $h$ front and the rise of $a$ at a cell.

Of the 640,000 parameter sets scanned, 137,235 had $A_a \geq .569$, where we anticipate that $a$ activation can be transient. Any parameter sets in which $a$ was transiently high, but fell back to a value near the low fixed point (an impermanent front) were in this set. Additionally, persistent activation of any sort was exceedingly rare for these parameter sets, confined to those where $A_a$ was very close to the cutoff of .569 or $u$ was never produced in significant amounts. Most of the parameter sets (≈95%) with the offending values of $A_a$ can be described as stalled solutions, given reasonable initial conditions. The balance show complicated dynamic behavior.

Of the remaining 502,765 parameter sets, we predicted 241,572 (or 48%) would have no propagating solution. Of these, 208,348 (86%) were unequivocally stalled. About three-quarters of the remaining 33,224 displayed some sort of fairly well behaved propagating solution. Most of these parameter sets gave patterns that either had multiple adjacent active cells, or propagated very quickly, in clear violation behind our analytic predictions. The remainder showed behavior suggesting pathologies in the numerical integration itself. We examined a subset of these pathological cases individually, pursuing them with tighter error tolerances. Subjected to this treatment, they resolved cleanly into one of the well-behaved classifications. We should mention that a prediction of a propagating solution does not necessarily imply that other attractors cannot exist. Indeed, there is always an attractor representing propagation failure for a sparse-enough initial pattern. Less universally, there can be an attractor representing a fast-propagating front with no patterning if the maximum source-density is high enough to push the important $h$ dynamics faster than $a$. What other asymptotic solutions might exist between these two extremes, and what transients are involved in approaching them, are interesting questions, though analyzing them in this 1D system is unlikely to yield much in the way of biologically relevant information: We are mainly concerned with the well-behaved patterning solutions.

Our theory predicts that the other 261,192 parameters sets have some sort of propagating solution that can be understood within our theoretical framework, provided the slow-$h$ assumption holds. Of these, 89% had an easily-classified propagating solution, and about 1% seemed truly stalled. The remainder seemed to yield easy-to-classify behavior only when given "special treatment," i.e. integrated with tighter tolerances, for longer times, and over larger domains. For 24,213 parameter sets the predicted self-consistent pattern was a front of uniform activation. This behavior, which must be regarded separately from cases where a uniform front was observed in defiance of expectations, was observed in 19,855 of the 24,213 cases (82%), with the remainder showing more exotic behavior. These parameter sets, with their dense patterns and minimal inhibition, tend to produce large amounts of activator and very fast fronts, and strain the assumptions of the model, but it is unclear what qualitative distinction can be drawn between slow-$h$ and fast-$h$ behavior. We pursued quantitative pattern analysis on those 236,932 parameter sets where there was a predicted pattern, other than uniformly high $a$.

It should be clear that our predictions about the behavior of parameter sets are, in broad, qualitative terms, correct. In the cases of the bad predictions one should keep in mind that our parameter space search cast a very wide net. The founder parameter set is a solid, well-behaved citizen, but the parameter space we explored extends past parameter regimes which are physiologically meaningful, deep into regimes which we now know are almost ridiculous.

*5.3.2. Quantitative results*

We now turn our attention to the analysis of the simulation data from parameter sets where we predicted a pattern-forming, propagating front. It is important to stratify these parameter sets by the degree to which they meet our assumptions. The first assumption, and the easiest to apply, is that up-means-up and down-means-down: The production of $h$ and $u$ by cells at the low steady state must be negligible, and the production of $h$ and $u$ at the high steady state must be significant. The high steady state assumptions for $h$ and $u$ have already been applied since these production rates were needed in calculating self-consistent pattern-forming solutions. The low steady state $h$ and $u$ production, however, we simply assumed to be zero in the analysis. This is a good assumption for the fly system, but it was occasionally violated by randomly chosen parameter sets. A strong low-steady-state criterion for $u$ is that at the "point of no return," i.e. the unstable steady state of $a$ at zero activation, the amount of $u$ produced is less than half what would be required for the producing cell to completely inhibit itself. A reasonable low-steady-state requirement for $h$ is that the equilibrium amount of $h$ produced when every cell is held at the unstable steady state is less than would be required to activate an uninhibited cell. Of the 236,926 parameter sets remaining after the previous exclusions, 151,450 simultaneously meet the high-low criteria, and thus constitute the parameter sets which test in detail our analytic treatment of the model equations. The specific analytic predictions we made for these sets break down as follows: 136,620 have a single propagating regularly spaced solution, 537 have multiple propagating regularly spaced solutions, and 14,293 have no period-1 self-consistent propagating solution, but may have solutions where the spacing between active cells varies between 2 values; we chose not to explore such possible solutions further.

76,118 (56%) of the 136,620 parameter sets predicted to have a single, attractive patterning solution made a propagating pattern with single, isolated activated points. Of these, we predicted the correct spatial period, $q$, for the solution in 97.1%. This degree of agreement is striking, as shown in Fig. 9, and is much better than that achieved using the simpler approximation in which the $h$ front is given a step function profile, which is only 62.1% accurate, and clearly systematically biased as shown in Fig. 10. Its bias towards predicting a period that is too short is not surprising: the step function model assumes that the first cell that could possibly be activated is the one that is actually activated, even when it sees a much higher $u_{ns}$ than subsequent cells. It remained to check whether violation of the slow-$h$ assumption could account for the 44% of parameter sets that did not violate any of the criteria already applied but that nonetheless did not agree with our analytic prediction.

In general, these parameter sets yielded behavior in which multiple adjacent cells were activated in the final pattern. The actual behavior in these cases ranged from unpatterned propagating fronts, in which every cell was activated (recall that some of the parameter sets that were predicted to stall also showed this behavior), to complex patterns of activated cells not obeying any obvious periodicity, to regular-appearing patterns of multiple active cells separated by multiple inactive cells. While measures of the

"average" expressed pattern period and spacing (in non-uniform solutions) showed significant correlation with the predictions, the absolute accuracy of the predictions was much lower than for parameters that showed basic OUID patterning. Additionally, it is unclear what such observations mean given the qualitative diversity of this group.

In qualitative classification of patterning behaviors discussed so far, we have not yet attempted to evaluate quantitatively how well the assumption of slow $h$ dynamics is met for different parameter sets. The specific time scales requiring comparison are: A) the time $T_a$ it takes a recently activated cell to reach the $a$ level necessary to inhibit its nearest neighbor, and B) the time it takes the propagating front of $h$ to progress 1 lattice site. These timescales are not strictly independent, and evaluating them separately, without careful integration of the full model, requires further approximation. We use the self-consistently calculated front velocity to estimate the second timescale, which we take to be of order $1/v$. The first timescale we approximate as the time it takes an isolated, uninhibited cell to progress from the steady state $a$ at bifurcation, to the level where its nearest neighbors are completely inhibited. This value has a non-trivial dependence on $h$ dynamics, but approaches a well-defined lower limit corresponding to constant, maximal activation: $\bar{g}(h)=1$. The physical assumption at work here is that the rise time of a cell depends only weakly on $g(h,u)$, because during much of the time the $a$ level is increasing, enough self-$u$ is produced that $g(h,u)$ is essentially zero, whatever value $h$ takes.

A scatter plot of parameter sets on axes reflecting these two time scales shows a clear separation between simple OUID patterns and more complicated cases (Figs. 11 and 12). To determine more precisely how well the time scales $T_a$ and $1/v$ alone predict a parameter set's behavior, we looked for the line in the $\ln(T_a)$, $\ln(1/v)$ plane that best separates parameters yielding patterns where only isolated activated cells appear from those for which every cell is activated by the passing front (Appendix C). This line is shown in Figs. 11 and 12; it discriminates between these two cases with a sensitivity, specificity, positive predictive value, and negative predictive value all substantially above 90%, indicating that a parameter set's qualitative behavior is indeed largely determined by the timescales of front motion and of cell activation [52].

The classifier just described ignores the minority of solutions that yield patterns containing both inactive cells and blocks of more than one adjacent active cells. As Fig. 12 shows, parameter sets showing such behavior cluster around our separating line, in the region of parameter space in between the more easily classified patterns. This suggests that as one varies parameters so as to reduce the timescale separation between $a$ and $h$, one will go from a situation where a propagating solution with isolated activated points is supported, to a region where only more complex patterns are supported, to, finally, a limit where the only propagating solution is one for which all the cells end up activated. The exact nature of this transitional region, unlike that of the zone between OUID patterns of different periodicity described in Sec. 4, is likely quite complex. It is also difficult to study, in part because of the very slow transients that can occur before the truly asymptotic long-time pattern appears. It is clear, however, that the transitional region holds parameter sets that can produce not only relatively exotic patterns, but also multiple patterns for a single parameter set as initial conditions are varied. In essence this multistability stems from the fact that the front speed depends not only on parameter values, but on the density of active cells (and thus of $h$ sources) in the pattern; it is possible, for one choice of parameters, to have relatively good separation of timescales when the density is low but to lose the timescale separation completely when the system is initiated with a high active cell density. It is finally worth noting that a similar transition (with similar complexities in the transition region) is found as $h$ is sped up between parameter regimes where no propagating fronts are possible and those where rapidly propagating, poorly patterned fronts occur.

The analytic theory of Sec. 4 predicts not only the spatial pattern period, but also the front speed. We would expect that the quality of these predictions should increase with longer times between the activation of cells, since the (comparatively) invariant activation time of a single cell, which our calculations effectively set to zero, will have a relatively smaller effect on the overall front speed under these conditions. For the parameter sets where we correctly predict the presence of a OUID pattern and its period, this is the case, as shown in Fig. 13.

## 6. Discussion

Although they have now been the subject of serious study for decades, activator-inhibitor systems continue to demonstrate new and unexpected behavior. Here, we have shown how coupling a simple activator-inhibitor subsystem to a longer-ranged diffusible activator can lead to front-driven pattern formation by a novel switch and template mechanism. The defining feature of this mechanism is its reliance on bistable switches which are flipped from a low to a high state in certain cells based on inputs from longer-ranged diffusible signals. Such behavior appears naturally in models in which cells are treated as discrete objects and certain genes self-activate cell-autonomously, with the concentrations of the self-activating species in one cell not depending directly on their concentrations in adjacent cells. Our dissection of the simplest, one-dimensional version of switch and template pattern formation has emphasized the essential role of a separation of timescales between the activator-inhibitor subsystem and the longer-ranged activator that drives front motion. Specifically, we have demonstrated that our model can be solved analytically in the limit that the former is much faster than the latter and that our solution in this limit correctly predicts the behavior of the full model for a substantial range of parameter values. As one might expect, however, our analytic predictions begin to fail as the two timescales approach each other; simple patterns built up from repeating units containing only a single active cell can then give way to far more intricate behavior. The exact structure of this boundary region remains obscure, but our analysis does make predictions about the qualitative arrangement of solutions in parameter space; these are summarized in the schematic bifurcation diagram of Fig. 14. Our basic insights from the one-dimensional model are directly applicable to the more complex and biologically relevant two-dimensional case. They both strongly suggest that switch and template pattern formation can operate robustly in biological systems and constrain the parameter regimes where this operation can occur.

Although many of the features found in our model have appeared individually in previous models, the consequences of coupling them together have not previously been described. Starting with Turing, one major theme in the study of reaction-diffusion systems has been the possibility of steady states that are unstable to finite-wavelength perturbations [1, 53, 54]. While the continuum has been examined most extensively, there have also been many studies that have concentrated on discretized systems where isolated cells become active [55-57]. In particular, more than one system has been described in which a patterned field expands into a region in an unstable state; in this case, the linear instability of the unstable state largely determines what final pattern is selected [58, 59]. The patterning system we have discussed here, in contrast, does not have a finite-wavelength linear instability. In its reliance instead on a bistable activator-inhibitor subsystem, our mechanism bears some resemblance to the formation of domain patterns in the Fitzhugh-Nagumo and related models [16-18], but the fact that auto-activation is strictly cell-autonomous leads to much more pronounced multistability among different patterns, while the presence of the long-ranged activator allows for front-driven pattern formation of a sort not usually associated with domain patterns. This prominence of lattice effects and front motion is reminiscent of work on front stalling in discrete systems [47, 48, 60] and of the well-known clock and wavefront mechanism [4, 61]. Our model differs from these, respectively, in its ability to generate stationary spatial patterns and in the absence of any oscillations. Importantly, in classic clock-and-wavefront patterning, the spatial period depends both on the frequency of the cell-autonomous oscillators and on the speed of front propagation, whereas in our system it is set directly by the range of the inhibitory signal and is largely independent of the details of the dynamics of other components of the system.

Although the basic model we have studied, Eq. (4), is a set of ordinary differential equations, our ultimate understanding of its behavior is more akin to what one might expect for a finite state machine or a Boolean model. This is significant on at least two counts. First, it reinforces the growing evidence that switch-like behavior plays a major role in fate specification during development [62-65]. Indeed, the fact that our model robustly engages in patterning associated with its limiting behavior as an array of switches, and that it moreover does so for parameter values consistent with the observed physiology of the fly eye imaginal disc, suggests why evolution might favor such a pattern formation solution. Second, our ability to pass from a continuous differential equation model to a hybrid object with a more discrete flavor gives an intriguing hint of how one might begin to analyze more complicated developmental models, involving multiple interacting genetic circuits. In such situations, an ability to coarse-grain the initial, detailed network model is essential; one way to do this is to identify functional modules consisting of several genes and to replace them with a single coarse-grained circuit element. Here, we have carried out just such a program for a simple model and have shown that it leads to robust pattern formation through a novel switch and template mechanism.

## Acknowledgements

We are grateful to Nick Baker and Boris Shraiman for ongoing collaborations related to this work. This research was funded by the NIH under grant GM047892.

## Appendix A: Numerical Procedures

Each of the 640,000 random parameter sets (Sec. 5.1) was subjected to several numerical and analytical tests. First, as an undirected exploration of the system's behavior, Eq. (5) was solved for each parameter set on an array of 1024 cells with periodic boundary conditions. All three fields ($a$, $h$, and $u$) were initially set to, except for $a$ on 100 adjacent cells where $a$ was chosen randomly and independently for each cell from a uniform distribution on [0, 0.25]. The equations were then integrated forward in time using an Euler integrator that treated the diffusive interaction terms fully implicitly. Each model was integrated forward in time 5000 time steps with $dt = .06$. All of the basic behaviors discussed in the paper (non-patterning fronts, stalled patterns, patterning fronts, fronts producing complicated patterns and transient activation—see Sec. 5.2) were observed in this test. Patterns were analyzed by eye to get a sense for the scope of the problem, and algorithmically to systematically classify the results. Another, similar test was conducted on a subset of stalled fronts using random uniform variants up to 0.35 instead of 0.25 for the initial $a$ values on 100 cells. With these initial conditions, some of the stalled solutions became moving-front solutions, demonstrating this simple predicted initial condition dependence.

The first step in automatically classifying patterns was to apply a threshold to $a$ corresponding to halfway between the zero-activation high steady state and the zero-activation intermediate unstable steady state (the "point of no return"). Cells with $a$ above this threshold were considered active. For parameter sets where these steady states do not exist ($A_a > .569$) an arbitrary threshold of 0.5 was used. The easiest behavior to classify, in general, is transient activation, as it requires only that one see a point that was once above threshold go below threshold. It is easy to classify the non-patterning fronts next. Because the range of the inhibitor is typically short, we decided to classify as non-patterning any front that showed at least 20 consecutive cells above threshold behind the most recently activated cell at the end of 5000 time steps. If the front overran the entire 1024 cell field in the allotted integration time, the front was additionally classified as "fast," and the last saved time point where the front had not yet crossed the entire field was used to evaluate the pattern. The vast majority of "fast" fronts were unpatterned, but there were exceptions. For a parameter set to be considered regularly patterning, the most recently created five groups of adjacent active cells had to consist of single active cells, and 3 of the 4 intervening gaps had to be equal in size. The solutions producing complicated patterns were subdivided into those with multiple adjacent cells in one of the most recent 5 groups, and those without. The first group dominated this category. To be considered stalled, a front had to produce no new active cells between time steps 2500 and 5000. Slipping through the cracks in this analysis are parameter sets that form very slowly propagating fronts. Indeed, parameter sets not conforming to any of the descriptions above were provisionally labeled "unknown behavior," but upon detailed examination most proved to produce solutions that activated <5 cells, but did activate at least 1 in the interval $2500dt - 5000dt$, thus failing the test for being stalled.

To compare these numerical results with the analytic theory of sec. 4, we followed the steps outlined in that section. We needed to calculate the amount of inhibitor at the points ahead of a patterned halfspace (which simply requires summing a geometric series) and the time when $h_{crit}(u_x)$ is exceeded for each of these points, which entails solving for the self-consistent velocity of the pattern, $v_q$. Once that is calculated, the priority of the point representing continued patterning must be established by calculating $h$ and $h_{crit}(u_x)$ at its neighbors. $h_{crit}(u_x)$ is easily calculated by setting $\partial_t a$ to zero, finding $g(h)$ as the root of the resulting polynomial, and then inverting that function if it is less than 1. $h_x(t)$ was constructed numerically, and a standard root-finding algorithm was used to solve the relationship $h_x(t) - h_{crit}(u_x) = 0$ for $t$ at all integer $x$ up to the maximum value of $x$ where $h_x(\infty) - h_{crit}(u_x) > 0$. The numerical approximation for $h_x(t)$ involved summing contributions from more and more distant active patterned sites according to Eq. (14), using a Runge-Kutta integrator with adaptive step size (because of the presence of more than one time-scale in the integrand) for the time integral, until one of two truncation conditions was met. The first truncation criterion was rarely used and involved a simple truncation if the contribution from the last patterned site was less than $10^{-11}$ of the running total. The second truncation method involved evaluated the ratios of contributions of consecutive sites, and, in the event the relative change in these became less than .01, extrapolating the further contributions as the total of an infinite geometric series with the appropriate decay constant, which gives excellent results.

With this new information, a second pass over the parameter sets was made, setting initial conditions and integration parameters according to the predicted patterning behavior. The initial conditions for all cells and all fields were zero, except for one cell at the end of the (no longer periodic) array which had $a$ at the high steady state. $h$ was put into the system as a time-dependent boundary condition based on the solution to the unpatterned continuum problem with the appropriate constants, and corrected to account for the $h$ produced by the initial 1-cell prepattern. The time-step, $dt$, was set to be .02 times the amount of time the front was expected to take to propagate 1 lattice unit, or .06, whichever was smaller, and the equations were integrated for twice as long as we anticipated it would take to produce 5 active cells. This led to some very long integrations. The time of each cell's activation was recorded and used to calculate the front speed. Pattern classification was conducted by methods similar to those described above. The main differences in the classification between these parameter sets were that some parameter sets that yielded non-patterning fronts originally yielded patterning ones, and those that were too slow to classify in the previous test were shown to propagate and pattern as expected.

## Appendix B: The Interaction of Front and Template

We consider a uniformly translating front of $h$ interacting with an inhibitor template exponentially decaying in space, $u$.

$$
\begin{aligned}
h_x(t) &= h(z) \\
z &= x - vt \\
u_x &= u_0 \lambda^x
\end{aligned}
\tag{30}
$$

The critical value of $h$, $h_{crit}$, at which the switch in each cell is flipped from low to high depends on $u$:

$$
\begin{aligned}
h_{crit}(u_x) &= H \left[ \frac{1 - h_c - h_c \left( u_x/U \right)^{m_u}}{h_c \left( 1 + \left( u_x/U \right)^{m_u} \right)} \right]^{-1/m_h} \\
h_c &= h_{crit}(0)
\end{aligned}
\tag{31}
$$

Here, we have introduced the variable $h_c$ to denote the critical value with no inhibitor. The continuum approximation for $h(z)$ was given in Sec. 4.2.1 and is

$$h(z) = s_0 \begin{cases} 1 - \left(\frac{v+c_1}{2c_1}\right) e^{\frac{-v+c_1}{2D_h}(z)}, & z < 0. \\ \left(\frac{-v+c_1}{2c_1}\right) e^{\frac{-v-c_1}{2D_h}(z)}, & z \geq 0 \end{cases} \qquad (32)$$

$$c_1 = \sqrt{v^2 + 4D_h}$$

Since the function is actually only sampled at integer $x$, we expect the first cell where $h_x(t)$ exceeds $h_{crit}(u_x)$ as time goes forward to be one of the integers flanking the value of $x$ at which the continuum $h(z)$ first surpasses the continuum $h_{crit}(u_x)$. At this first crossing of the two curves, both the functions themselves and their tangents must coincide:

$$\begin{aligned} h(x - vt) &= h_{crit}(u_x) \\ \partial_x h(x - vt) &= \partial_x h_{crit}(u_x) \end{aligned} \qquad (33)$$

We want to know two things about this point of intersection. First, how sensitive is it to changes in template patterns given a particular $h(z)$? If it changes by less than one, then only stable OUID patterns, or high-period patterns as discussed in Sec. 4.2.2 can exist. Second, sensitive is it to changes in $h(z)$, as if, for instance, some random errors had occurred in the templating process? If it changes by very much less than 1 for the $h(z)$ that would be produced by patterns that differ in wavelength by one, then having more than one stable OUID pattern supported by the same parameter set will be proportionally unlikely.

We proceed by expanding $h_{crit}(u_x)$ and $h(z)$ in Taylor series up to second order in $x$ about their point of most likely contact, i.e. where the following relationships are satisfied:

$$u_x = U \\ z = 0 \tag{34}$$

This gives the following formulae:

$$h_{crit}(x) \approx H\left(\frac{2h_c}{1-2h_c}\right)^{\frac{1}{m_h}} - \\ H\left(\frac{2h_c}{1-2h_c}\right)^{\frac{1}{m_h}}\left(\frac{m_u \log[\lambda]}{-2m_h + 4h_c m_h}\right)\left(x - \frac{\log[U] - \log[u_0]}{\log[\lambda]}\right) + \\ \frac{H}{2}\left(\frac{2h_c}{1-2h_c}\right)^{\frac{1}{m_h}}\left(\frac{(1+m_h)m_u \log[\lambda]}{-2m_h + 4h_c m_h}\right)^2\left(x - \frac{\log[U] - \log[u_0]}{\log[\lambda]}\right)^2 + \\ O\left(x - \frac{\log[U] - \log[u_0]}{\log[\lambda]}\right)^3 \tag{35}$$

and

$$h(x) \approx \frac{1}{2}s_0\left(1 - \frac{v}{c_1}\right) + \\ s_0 \frac{(v^2 - c_1^2)(x - vt)}{4c_1 D_h} + \\ s_0 \frac{(c_1 - v)(c_1 + v)^2(x - vt)^2}{16 c_1 D_h^2} + \\ O(x - vt)^3 \tag{36}$$

,

where we have opted to use the steeper branch of $h$, which was defined piecewise. The relative magnitude of the two second order terms, here, is significant, as we are interested in tangential contact. For small velocities, the magnitude of the second order term in the expression for $h(x)$ falls off as $1/D_h$ whereas the first order term falls off as $1/\sqrt{D_h}$, meaning that for higher $D_h$ the approximation becomes better, and the dependence on $x$ more linear, which is unsurprising. By comparison, the second order term for $h_{crit}(x)$ is dependent mainly on the steepness of the $u$ gradient, and the Hill coefficients $m_u$ and $m_h$. For typical parameters, this second order term is hundreds of times larger than that for $h(x)$.

We want to solve for tangential intersection of these two approximations, as mentioned. In general, the tangential intersection of two quadratics is easily calculated. For convenience, here, we apply the further simplification that $h(x)$ is linear,

$$h_{crit}(x) = r_2\left(x - \frac{\log[U] - \log[u_0]}{\log[\lambda]}\right) +$$
$$r_1\left(x - \frac{\log[U] - \log[u_0]}{\log[\lambda]}\right) + r_0, \qquad (37)$$

$$h(x) = w_1(x - vt) + w_0$$

where the $r$ and $w$ coefficients are the appropriate terms from the Taylor series. The linear approximation is unnecessary if one does not mind the cumbersome equations it produces. Solving for tangential intersection yields

$$x = \frac{-r_1 + w_1 + 2r_2\left(\frac{\log[U]-\log[u_0]}{\log[\lambda]}\right)}{2r_2}. \tag{38}$$

Taking the partial derivatives of $x$ with respect to $u_0$ and $s_0$ gives the sensitivity of this intersection point to, respectively, template source strength and pattern density,

$$\partial_{u_0} x = -\frac{1}{u_0 \log[\lambda]}$$

$$\partial_{s_0} x = \frac{(1-2h_c)^{1/m_h}\left(-c_1^2+v^2\right)(m_h-2h_c m_h)}{(2h_c)^{1/m_h} c_1 D_h H (1+m_h) m_u^2 \log[\lambda]^2} \tag{39}$$

The first derivative is, unsurprisingly, dependent on the source strength of the inhibitor and its length scale $u$. It is of order unity or less for typical parameters. The variation one can expect in $x$, then, from a template pattern of $q+1$, instead of the preferred period $q$ is given approximately by

$$\left|(\lambda^{q+1} - \lambda^q)\partial_{u_0} x\right| \ll 1. \tag{40}$$

The expression for the source-strength sensitivity is more complicated, but it is notable that there is a net factor of $\sqrt{D_h}$ in the denominator, as well as a net factor of $m_u^2 \log[\lambda]^2$, the first of which is high when $h$ is smooth, the second of which is high when the template is steep. This implies that for parameters that typify our assumptions, this derivative can be quite small. Multiplying it by a change in source density that is also much less than 1 (the maximum source density) suggests that the first point to be activated is relatively independent of small changes in source density, and thus, generically, only one patterning solution is supported in this limit, with two patterns supported infrequently, in proportion to the shift in $x$.

## Appendix C: Binary classification based on timescales

In this appendix, we describe how we obtained the line in Figs. 11 and 12 separating OUID patterns from patterns in which all cells are active; the basic motivation for finding such a separating line is discussed in sec. 5.3.2.

Given a line $\ln(1/v) = m \ln(T_a) + b$ in the $\ln(T_a)$, $\ln(1/v)$ plane, we can determine whether each point in Fig. 11 falls above or below that line. We already know whether the point corresponds to parameters that give an OUID or an all-up pattern. Based on these two binary decisions, each point can thus be assigned to one of four groups. Let $A$ be the number of points above the line with an OUID pattern, $B$ the number above the line but with an all-up pattern, $C$ the number below the line but with an OUID pattern, and $D$ the number below the line with an all-up pattern. If the line perfectly separated the OUID from the all-up patterns, we would have $B = C = 0$, and we can say that the line does a good job of classifying the patterns if $B$ and $C$ are small. More specifically, a good classifier should have a sensitivity $A/(A+C)$, specificity $D/(D+B)$, positive predictive value $A/(A+B)$, and negative predictive value $D/(D+C)$ all as close to unity as possible [52]. We thus defined the line that best separated the two sorts of patterns to be the one with the slope *m* and intercept *b* that maximized the product $A^2B^2/(A+B)(A+C)(D+B)(D+C)$ of these four measures. More standard choices, like the phi statistic $AD\text{-}BC/[(A+B)(A+C)(D+B)(D+C)]^{(1/2)}$, are also available, but we found that these performed slightly worse on our data (for which *A* and *D* are large and of the same order while *B* and *C* are small).

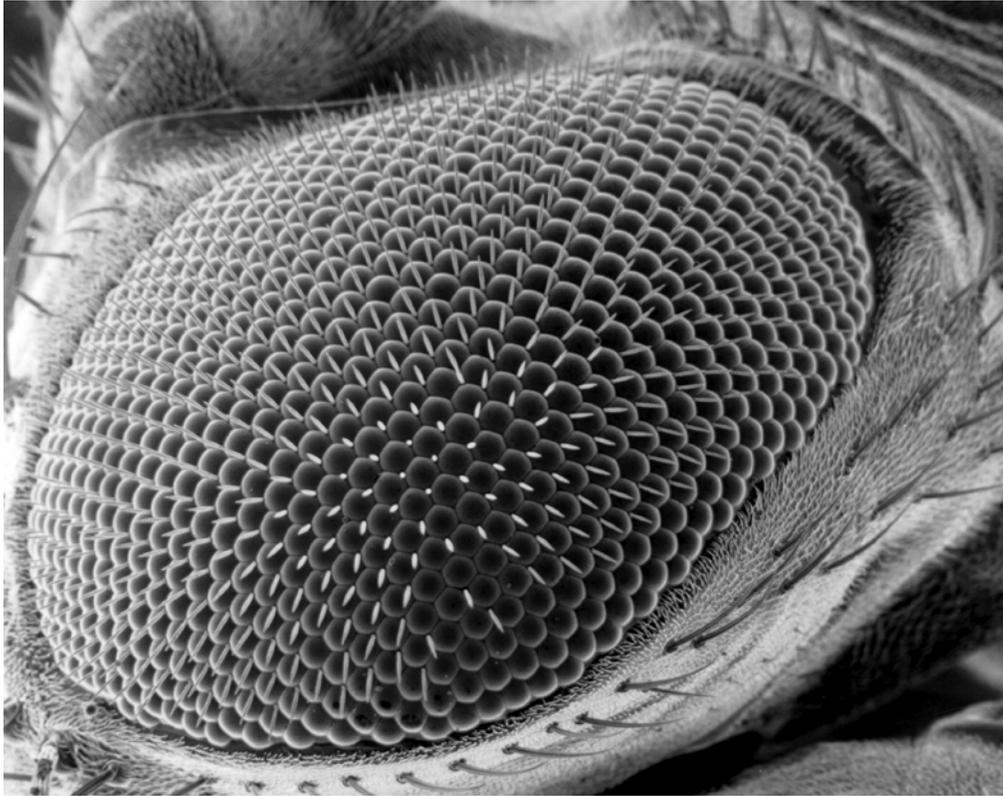

**Figure 1:** Scanning electron micrograph of the adult *Drosophila* eye. Each round facet is the lens of a photoreceptor cluster called an ommatidium. Each ommatidium is founded by a single photoreceptor neuron, the R8 cell, which is specified during larval development. The dramatic hexagonal order visible here is first observed in the spatial arrangement of these R8 cells. (Public domain image courtesy of Dartmouth Electron Microscope Facility.)

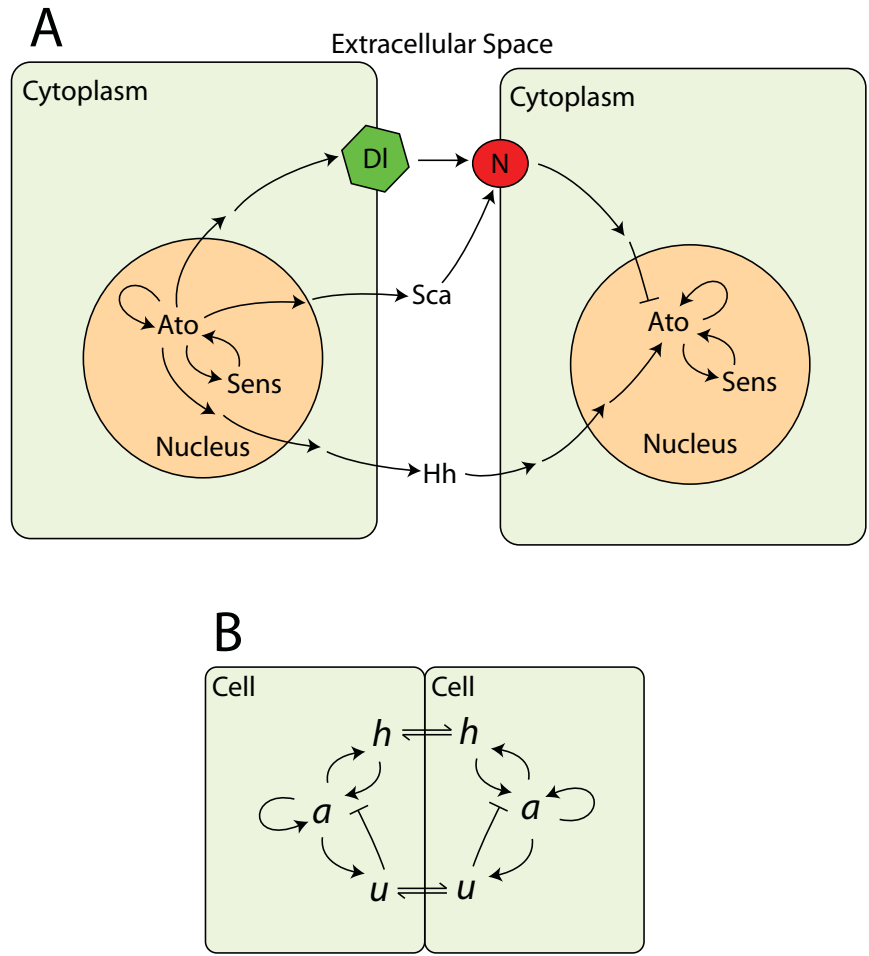

**Figure 2:** (Color online) Some interactions involved in patterning R8 photoreceptors in the *Drosophila* eye. (A) Intercellular and intracellular regulation involved in fate specification in the epithelium of the eye imaginal disc. Only signals originating at the left cell and being received by the right cell are shown, but all interactions may be assumed to be reciprocal. Pointed arrows show a positive, activating influence; blunt arrows signify inhibition. The locations of gene names reflect the subcellular localization of the gene product. This diagram is necessarily incomplete, and most of the signals transmitted really on other genes during their production and transduction. N, Notch; Dl, Delta, Hh, Hedgehog; Sca, Scabrous; Ato, Atonal; and Sens, Senseless. (B) The simplified model studied in this paper. Diffusible activation is represented by $h$, with inhibitory activity lumped into $u$. The variable $a$ takes the place of the proneural genes *ato* and *sens*. The multitude of different compartments present in a tissue are ignored, with each cell being treated as a lattice site, and intercellular signals moving on this lattice by diffusion (signified by the bi-directional arrows).

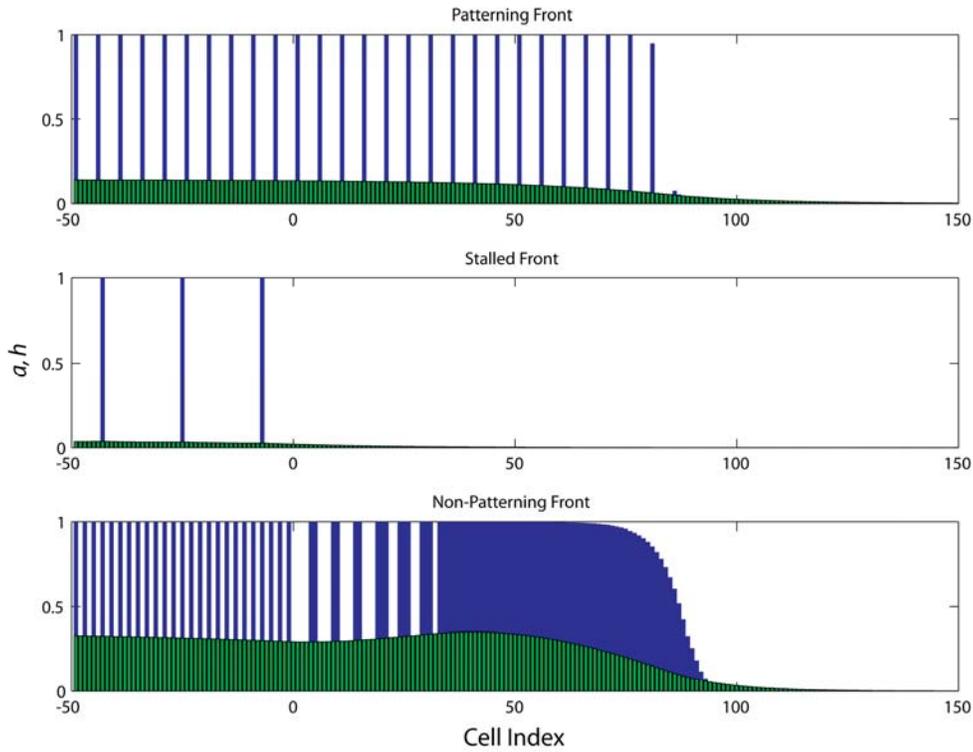

**Figure 3**: (Color online) Typical simulation results. In each plot, the activation of cells with negative indices was specified by initial conditions, but any cells activated with indices greater than zero represent propagation of a moving front. All simulations were conducted on lattices of 2048 cells by integrating Eq. (5) with the same set of parameters, varying only the wavelength of the initial pattern. **A**) A propagating front of $h$ (light, green) that produces a stable, regular pattern of $a$ (dark, blue). **B**) Propagation can fail if the $h$ produced by the initial localized pattern is insufficient to activate $a$ in additional cells. This always occurs for a sparse-enough prepattern. **C**) If the evolution of $a$ and $u$ is too slow for a recently activated cell to inhibit its neighbors before the $h$ front gets to them, a propagating, unpatterned front of activation may exist. This solution can exist for parameter sets that otherwise have only stalled solutions and for ones that also have patterning solutions. In this case it is induced by supplying a too dense prepattern.

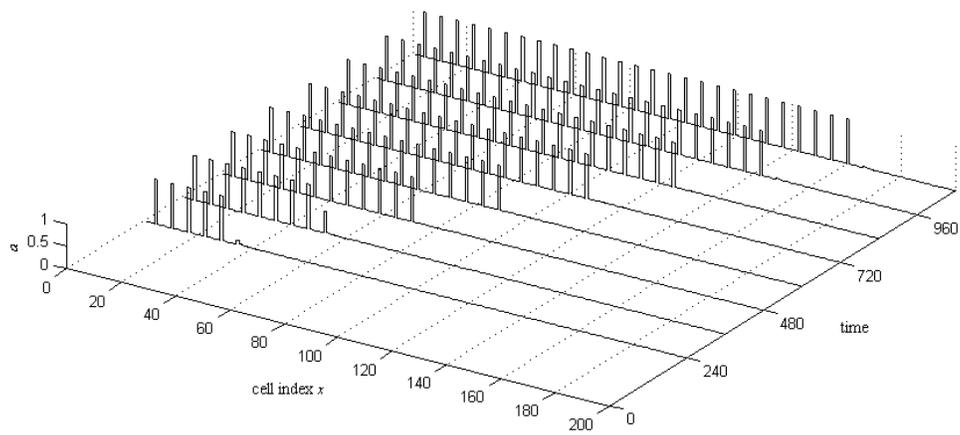

**Figure 4:** Spatiotemporal portrait of a patterning solution of our model, Eq. (5). The variable *a* is plotted as a function of spatial position (or cell index *x*) for 8 regularly spaced time points. As time progresses, the pattern expands with a constant speed while maintaining the same period. Each rectangular spike in the plot corresponds to a single activated cell.

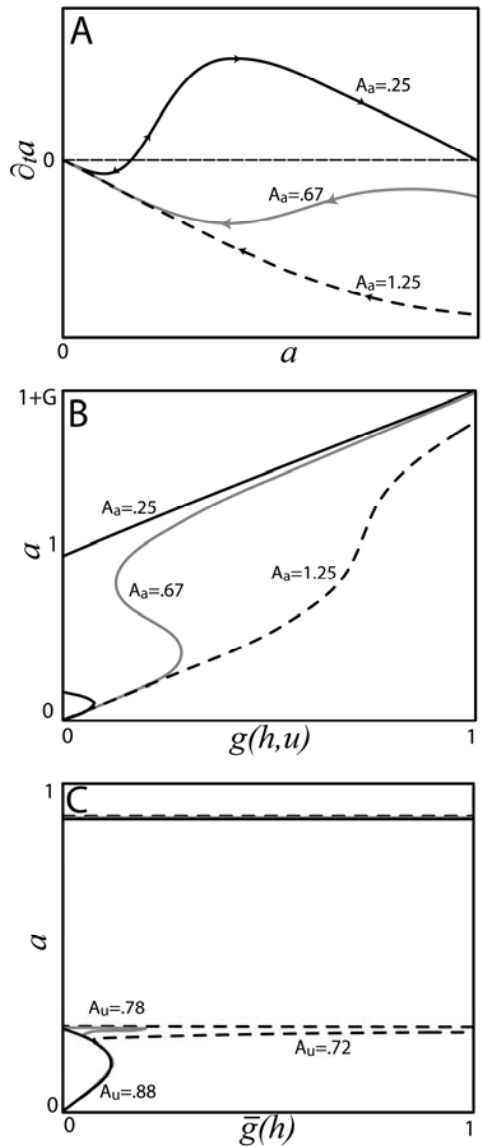

**Figure 5:** (A) $\partial a/\partial t$ versus $a$ for an isolated cell with $h=0$ or $u \gg U$. At low $A_a$ bistability exists even when $h=0$. At high $A_a$ there is no bistability. At intermediate values bistability can exist for some amount of external activation. (B) Steady states of $a$ as a function of the input from $h$ and $u$. (C) Steady states of $a$ including the effects of $u$ produced by the same cell, plotted as a function of the activating input $\bar{g}(h)$. The cell is assumed to receive negligible $u$ from other cells. As $A_u$ decreases, the high steady state

eventually becomes completely inaccessible to cells starting from $a=0$, even when $h$ is very large.

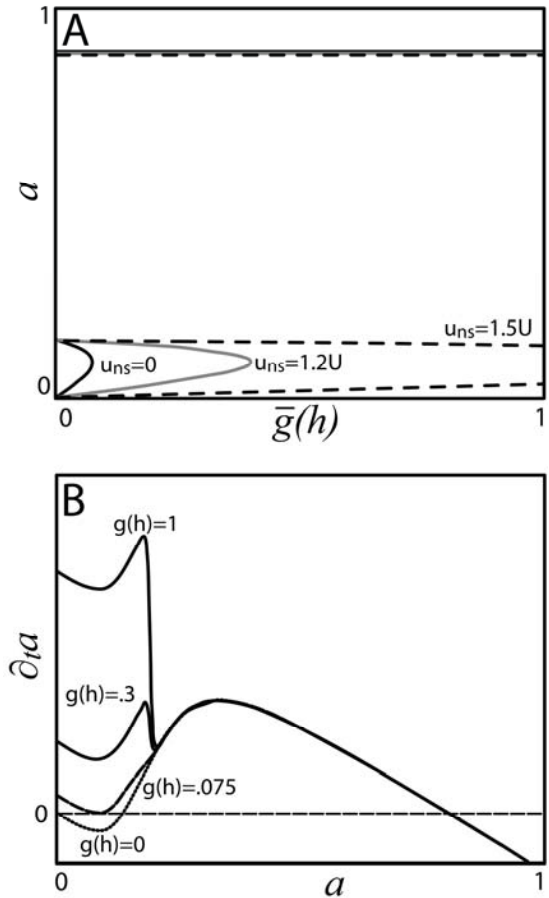

**Figure 6:** (A) The response of $a$'s bifurcation diagram to increasing amounts of externally generated $u$ ($u_{ns}$). The bifurcation value which represents loss of the low (stable) and middle (unstable) steady states proceeds from its unperturbed value, through higher values, to values that are unattainable with finite $h$. (B) The dynamics of $a$ with no external inhibition and various fixed activations. The effect of autoinhibition means that $\partial_t a$ is insensitive to $h$ for $a$ above a certain threshold, and thus that there is well-defined minimum amount of time between a cell's activation and its reaching the high $a$ state.

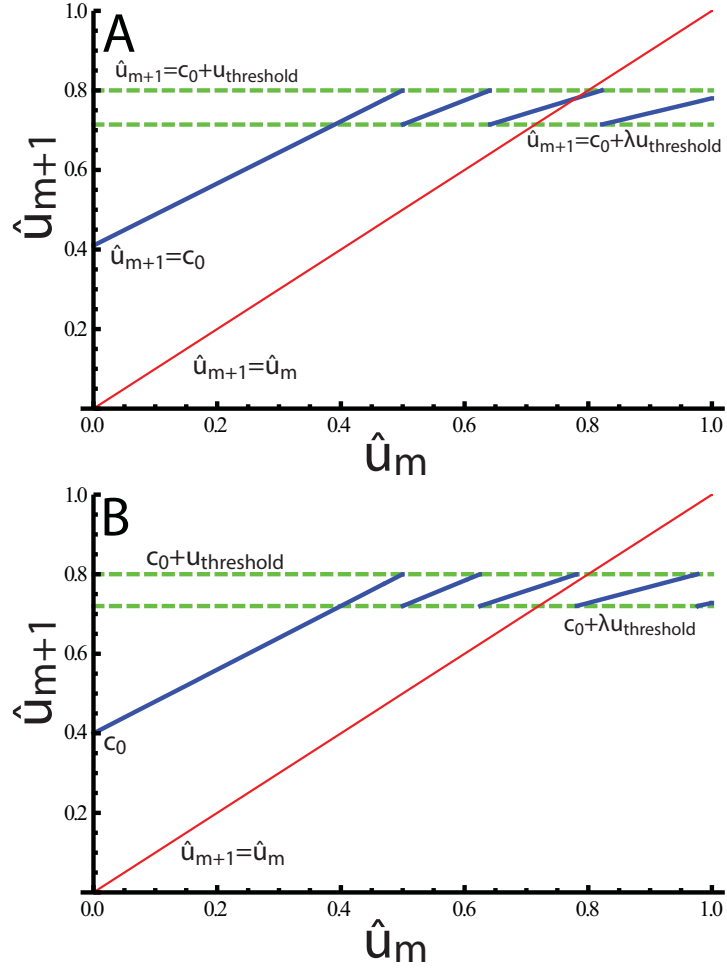

**Figure 7:** (Color online) The map relating $u$ at a newly patterned cell immediately after its activation to the amount of $u$ at the previous activated cell. The heavy, solid blue lines represent the map function, as given by Eq. (20). The top dashed line indicates the maximum amount of $u$ that still permits cell activation. The lower dashed line shows the minimum amount of $u$ at a point that also implies its neighboring point cannot be activated. In (A), the identity line (light solid line, in red) intersects the fourth line segment of the map function, implying the existence a single, stable 1-up-2-down pattern. In (B), the identity goes through a discontinuity, so that asymptotically the pattern will alternately have gaps of 2 or 3 cells between active cells.

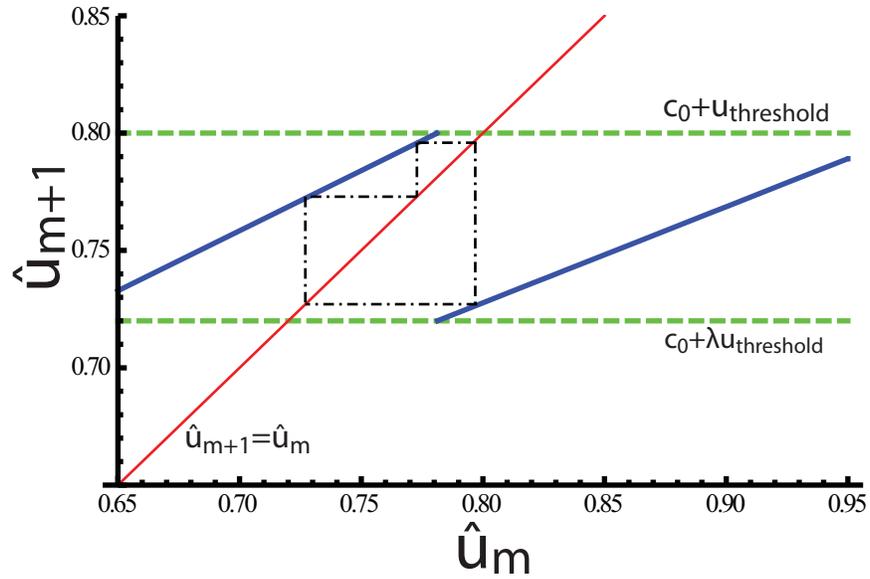

**Figure 8**: (Color online) A detail of the higher-period solution in Fig. 7(B). The attractive orbit of the map is shown as the dash-dotted line. In this case, the overall period is 10 cells, and consists of repetitions of the motif 1-up-2-down-1-up-2-down-1-up-3-down.

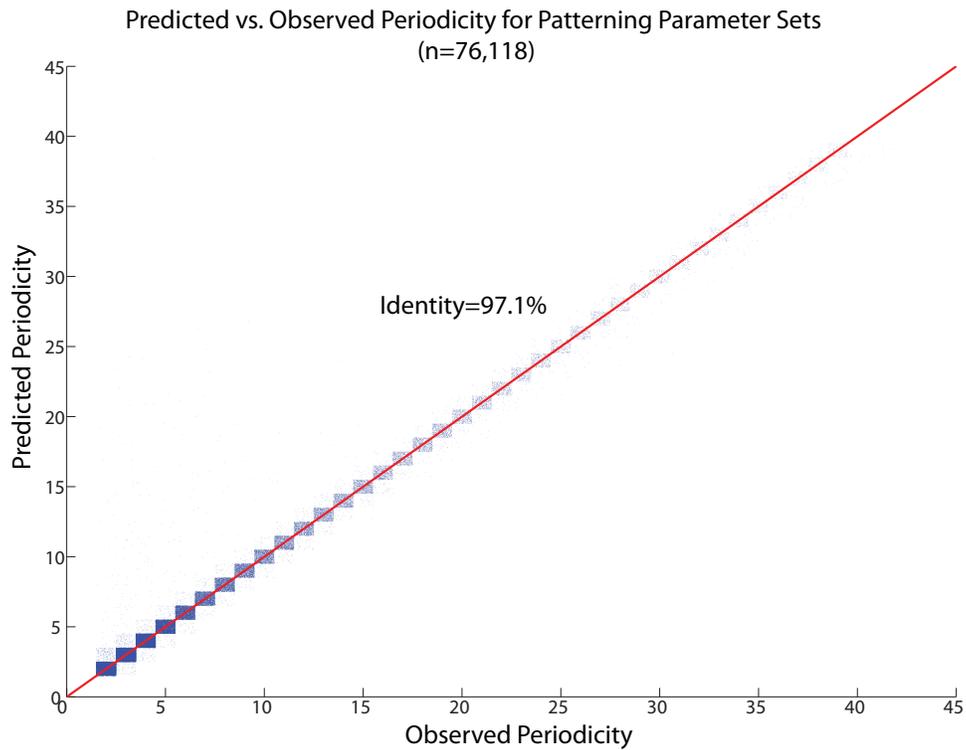

**Figure 9:** (Color online) Predicted vs. observed pattern period for parameter sets showing regular OUID patterns. Each blue dot represents a parameter set. The points described by a particular ordered pair of integers ( [*observed, predicted*], for instance [5,3] ) are assigned a random location within a square box of side 1 centered on those coordinates, to give an indication of the density of points. The points are densely concentrated along the identity line. More than 97% of parameter sets show perfect agreement.

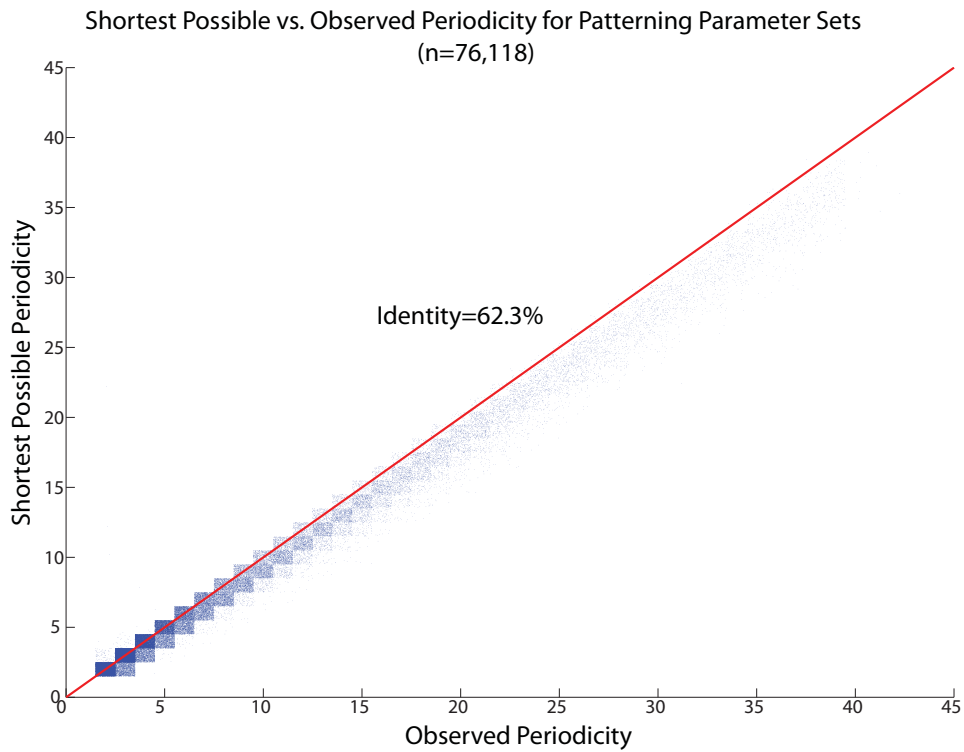

**Figure 10:** (Color online) Same Fig. 9, but with period predicted by the simpler step-function activator model of Sec. 4.2.2. The overall correlation of prediction and observation is still clear, but, as expected, there is a bias towards predicting periods that are too short.

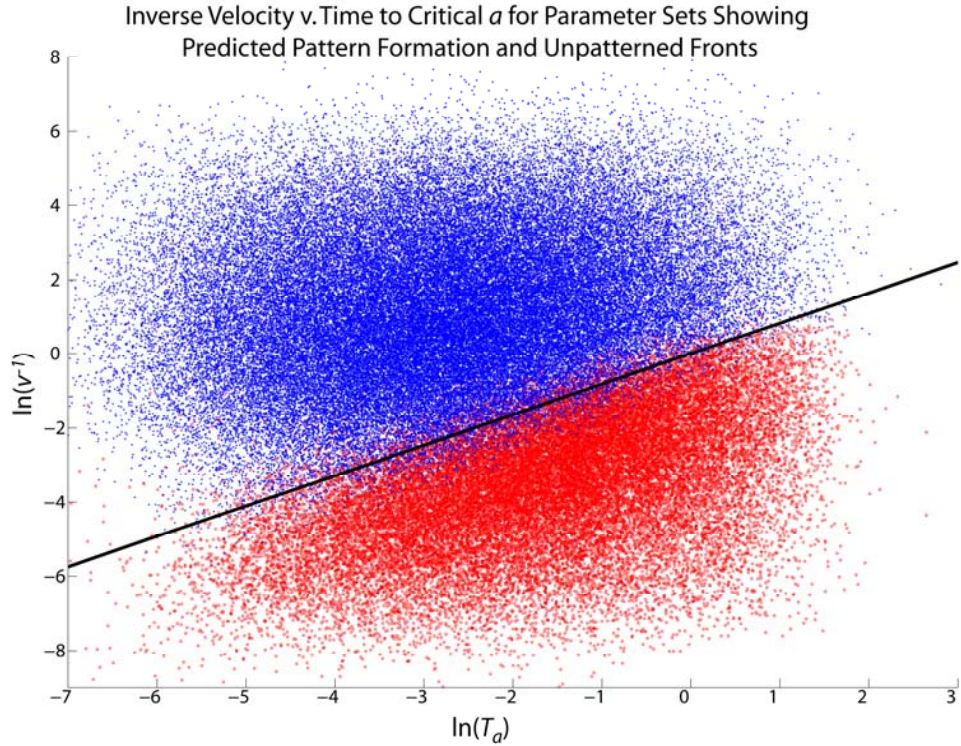

**Figure 11:** (Color online) Each point represents a parameter set for which pattern formation was predicted. The pattern was either observed as predicted (dark, blue), or an unpatterned (i.e. all cells active) propagating front was observed (light, red); parameter sets with other behaviors are not shown (see Fig. 12). Horizontal axis, shortest possible time for a cell experiencing maximum activation to reach high enough $a$ to fully inhibit its nearest neighbor. Vertical axis, inverse front velocity. This approximates the amount of time it takes the average $h$-front to advance one lattice site. The black line optimally separates the two possible outcomes. It successfully classifies about 95% of these parameter sets. The switch and template pattern formation mechanism begins to fail when the internal dynamics of a cell can no longer be considered fast compared to front propagation.

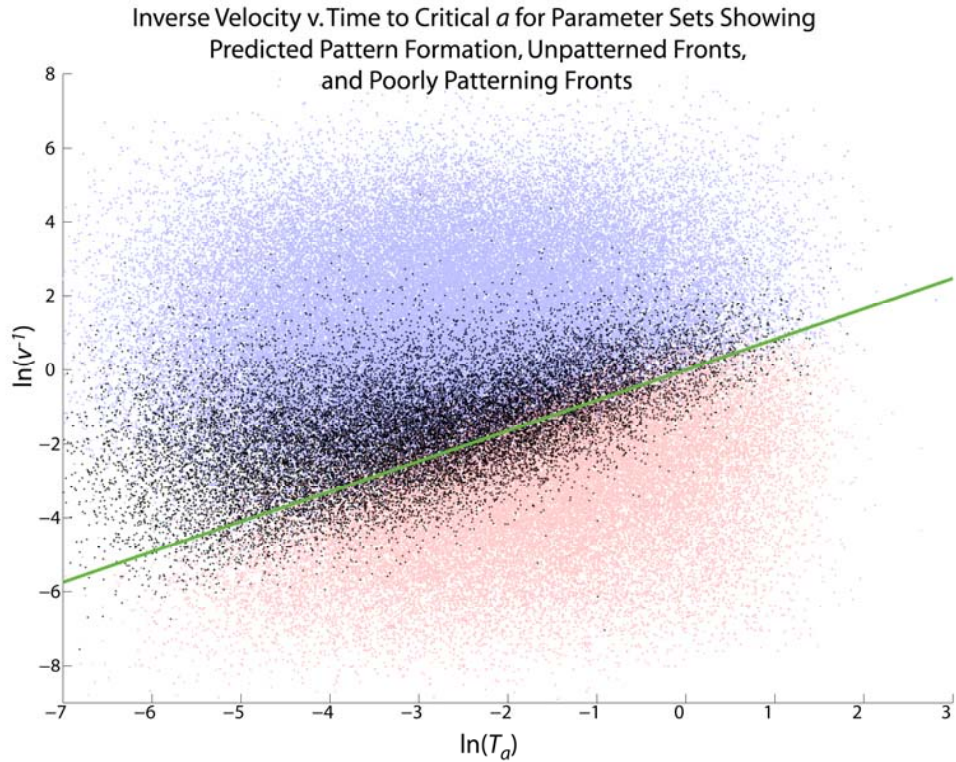

**Figure 12**: (Color online) Same as Fig. 11, but with the addition of the points for which a pattern was predicted, but neither that pattern nor a uniform propagating front was observed (black). Very complicated behavior was observed in this set, and these parameter sets are particularly prone to very long transients. Whether these solutions are in the process of settling down to one of the better-known behaviors (patterning or non-patterning) or are approaching other, more complicated limiting behavior is an open question. It is clear, however, that they tend to fall between parameters that lead to patterning and those that lead to uniform high $a$.

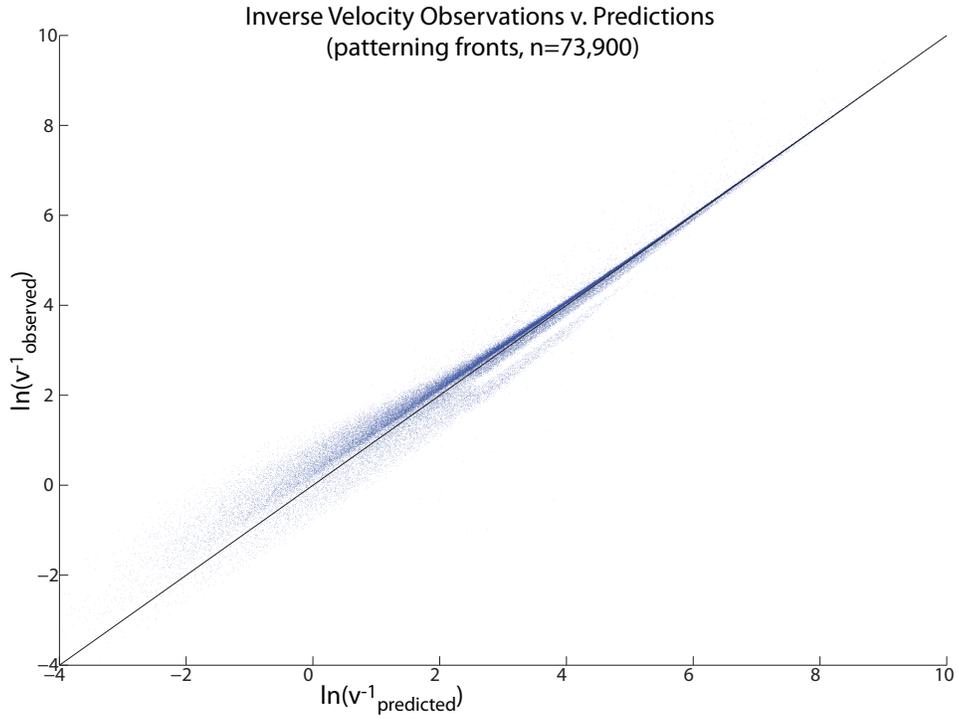

**Figure 13**: (Color online) Comparison of observed front velocity from integration of Eq. (5) to analytic predictions based on the fast-$a$ approximation, for parameter sets that lead to stably propagating patterns with the predicted wavelength. Each dot is a parameter set. The prediction becomes relatively better as the front slows down, as expected.

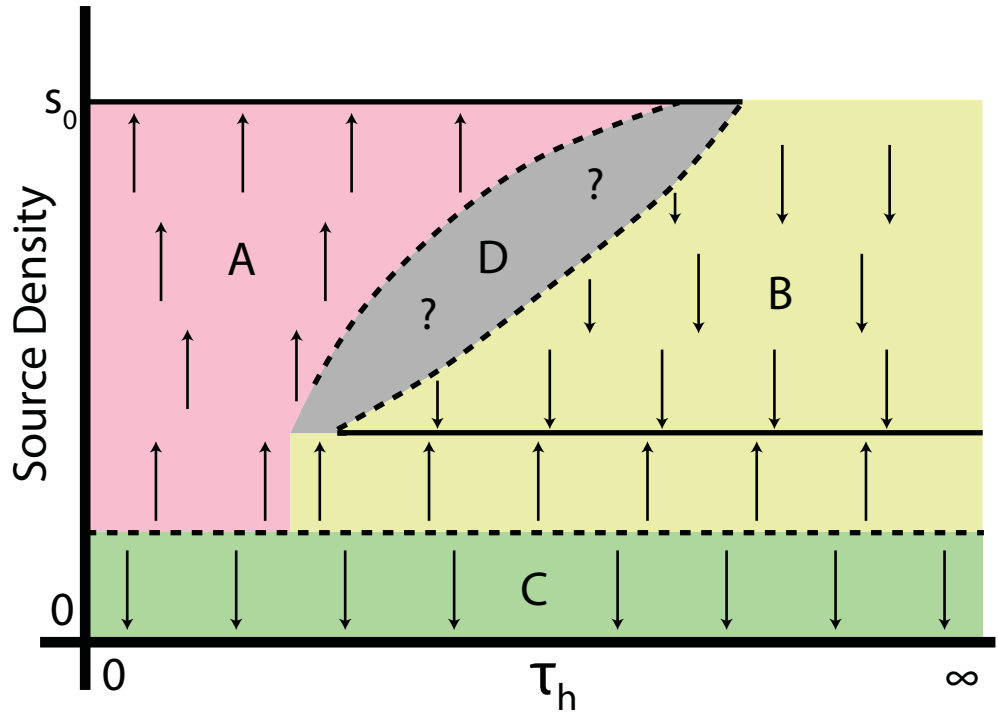

**Figure 14**: (Color online) Schematic one-parameter bifurcation diagram for front propagation and pattern formation, showing the stable solution types discussed in this paper and their basins of attraction. Solid lines, stable long-time behaviors; dotted lines, unstable behaviors. Arrows indicate the direction in which the system evolves over time. We take the source density for $h$ (i.e. the fraction of cells in an active, or high $a$ state) as the output state; it can range from 0 (stalled) to $s_0$, the maximum activity of a single cell. The bifurcation parameter $\tau_h$ controls the relative timescales of front motion and of activation of a single cell. Our analytic predictions (Sec. 4) apply for large $\tau_h$, and thus for regions B and C, which correspond, respectively, to stalled fronts and to OUID patterns. If the initial pattern density is too low, front propagation cannot occur, and the system lies in region C. A system that produces enough $h$, and where $h$ dynamics are slow enough compared to $a$ (region B), is attracted to a regular patterning solution. If $h$ dynamics are not slow, then many cells can be activated before any is able to inhibit

another, and a propagating front characterized by a maximum-density pattern is observed (region A). The structure of the boundary between regions A and B is unknown (gray, region D), but there are parameter sets where stable patterning and unpatterned front propagation are observed for different initial conditions (see Fig. 3).

| Parameter | Min/$p_{ref}$ | Max/$p_{ref}$ | Distribution |
|---|---|---|---|
| $A_a$ | .01 | 10 | Log |
| $G$ | .01 | 100 | Log |
| $H$ | .01 | 100 | Log |
| $m_h$ | .0625 | 1.25 | Linear |
| $U$ | .01 | 100 | Log |
| $\tau_h$ | .01 | 10 | Log |
| $A_h$ | .01 | 5 | Log |
| $D_h$ | .01 | 100 | Log |
| $A_u$ | .01 | 5 | Log |
| $D_u$ | .01 | 100 | Log |

**Table I:** Scanned parameters and ranges. Each model parameter was chosen randomly and independently from within a given range. The minimum and maximum values were set by the indicated ratios with the reference parameter set. For most parameters, the distribution that was sampled was $p\left[\ln\left(parameter/p_{ref}\right)\right] \propto constant$, the distribution identified in the table as "Log." The exponent $m_h$ was sampled uniformly over its range.